\documentclass{osa-article}

\journal{oe}

\articletype{Research Article}

\usepackage{lineno}
\usepackage{color}

\begin{document}

\title{Low coherence interferometric detection of the spectral dependence of the retro-reflection coefficient of an anti-reflective coated interface}

\author{Michel Lequime,\authormark{*} Imran Khan, Myriam Zerrad, and Claude Amra}

\address{Aix Marseille Univ, CNRS, Centrale Marseille, Institut Fresnel, Marseille, France}

\email{\authormark{*}michel.lequime@fresnel.fr} 



\begin{abstract}
The measurement of very low reflection coefficients of anti-reflective coated interfaces has become a key issue for the realization of precision instruments such as the giant interferometers used for the detection of gravitational waves. We propose in this paper a method, based on low coherence interferometry and balanced detection, which not only allows to obtain the spectral dependence of this reflection coefficient in amplitude and phase, with a sensitivity of the order of 0.1 ppm and a spectral resolution of 0.2 nm, but also to eliminate any spurious influence related to the possible presence of uncoated interfaces. This method also implements a data processing similar to that used in Fourier transform spectrometry. After establishing the formulas that control the accuracy and the signal-to-noise ratio of this method, we present the results that provide a complete demonstration of its successful operation in various experimental conditions.
\end{abstract}

\section{Introduction}
Anti-reflective coatings are undoubtedly one of the most important categories of optical interference coatings \cite{Macleod_2018,Raut_2011}. They are used in a wide range of applications, from consumer optics (photography, eyewear, LCD display) to high performance scientific instrumentation (Earth observation from space, interferometric detection of gravitational waves). Because of this broad range of applications, the reflection specifications of these coatings are extremely varied, ranging from percent or fraction of a percent over large spectral ranges \cite{Dobrowolski_1996,Lemarquis_2019} to a few ppm for specific wavelengths corresponding to laser emissions \cite{Okuda_2006}.

Here, we are interested in the latter, and the corresponding coatings are then often designated by the names V-coat or V-shape coatings, because the theoretical variation of their reflectivity in logarithmic units presents a V shape whose minimum is centered at the design wavelength. Such a AR coating response corresponds, for example, to the one that can be obtained by depositing on a substrate two layers of materials with high and low refractive indices respectively and whose thicknesses are adjusted in accordance to the value of their indices and the required central wavelength of operation \cite{Amra_2021}. Thus, to obtain a theoretical zero in reflection on a N-BK7 window at the wavelength of 1064 nm, using Niobium pentoxide (Nb$_2$O$_5$) as the high index material and silicon dioxide (SiO$_2$) as the low index material, we can use the following stack formula
\begin{center}
N-BK7 / 1.690H / 0.681L / Air
\end{center}
where H and L are quarter-wave thicknesses of high and low index materials respectively.

Independently of the manufacturing challenges that the reliable realization of such a deposition is likely to raise, even in the case of deposition machines using stable and energetic processes (Ion Beam Sputtering, Plasma Assisted Reactive Magnetron Sputtering) and high performance \textit{in situ} optical monitoring systems (see for instance \cite{Okuda_2006}), it is also a challenging task to accurately characterize the residual reflection value of the coated face with an accuracy of a few ppm and a perfect insensitivity to the optical properties (reflection, scattering) of the other face. The use of classical techniques \cite{Willey_2014}, such as the subtraction of the theoretical contribution from the rear face to the experimental reflection factor, the rough sanding of this rear face or its coating with a black paint, are not suitable here as they are adapted to cases where a precision of 0.1\% is sufficient. 

Two measurement methods that achieve the required sensitivity levels are the two-channel cavity ring down technique \cite{Cui_2017} and the use of a tunable laser with high side-mode suppression ratio \cite{Uehara_2004}. The former achieves sub-ppm accuracy, but is monochromatic (635 nm) and requires the use of a two-side coated component illuminated with a typical incidence angle of 5 degrees. The later gives access to the reflection spectrum of one of the faces, but again, it is necessary to illuminate the sample with a non-zero angle of incidence (2 degrees for instance), and the absolute calibration of the set-up with a ppm accuracy seems very difficult.

The method we describe in this paper provides an effective solution to address these challenges and allows to measure the reflection spectrum of the antireflection coated face of a plane window under normal incidence and with a much better sensitivity than ppm. Moreover, it allows to determine the spectral dependence of the phase shift to the reflection on this stack. Section \ref{sec:SetUpDescription} provides a description of the low-coherence, balanced-detection interferometric set-up used to record this reflected flux, while Section \ref{sec:DataProcessing} details the Fourier transform processing scheme implemented to extract the spectral dependence of this reflection coefficient, amplitude and phase. Section \ref{sec:Sensitivity} analyzes the theoretical SNR of this measurement method, as well as the influence on this SNR of the reduction of the width of the data processing windows, and the consequences on the result of this measurement of an angular misalignment of one of the faces of the sample. Section \ref{sec:ExperimentalDemonstration} describes the experimental results obtained on a 2 mm thick silica wafer with a V-shaped antireflection coating on one side, while Section \ref{sec:Discussion} provides a critical analysis of these results. Finally, Section \ref{sec:Conclusion} summarizes our main achievements and defines possible next steps in the development of this technique.

\section{Method}
\label{sec:Method}

\subsection{Set-up description}
\label{sec:SetUpDescription}

\begin{figure}[htbp]
\centering\includegraphics[width=0.80\textwidth]{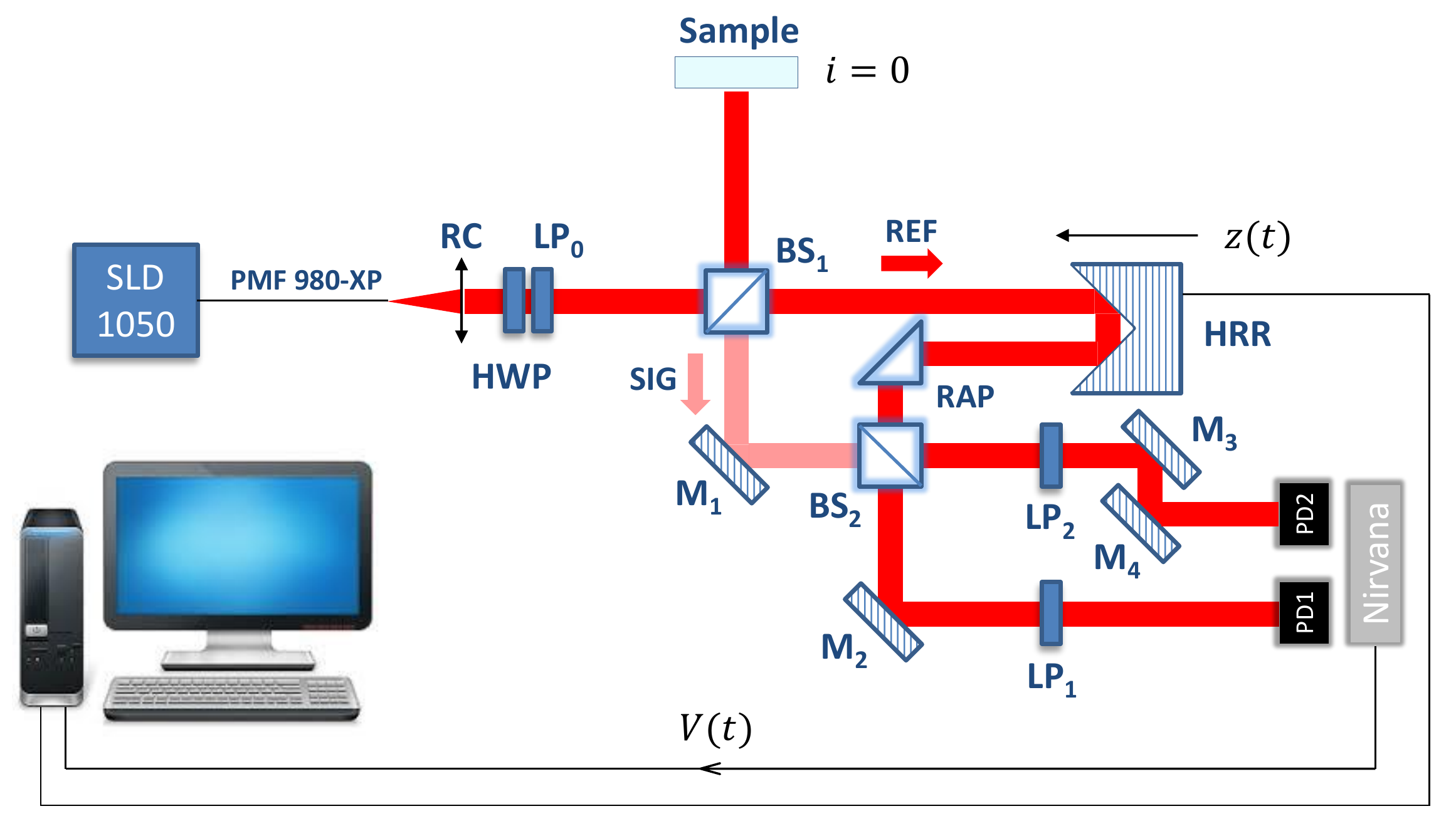}
\caption{Low-coherence balanced-detection interferometric set-up (BARRITON).}
\label{fig:BenchScheme}
\end{figure}

The set-up used to measure this reflection coefficient is referred to as BARRITON (for Back-scattering And Retro-Reflection by InterferomeTry with lOw cohereNce). It is an upgraded version of the one described in Reference \cite{Khan_2021} and is shown schematically in Fig. \ref{fig:BenchScheme}.

The linearly polarized light flux provided by a superluminescent diode (SLD 1050) centered around 1050 nm is coupled into a PM980-XP polarization-maintaining single-mode fiber (mode field diameter $2w_0=6.6$ $\mu$m @ 980 nm), whose output end is placed at the focus of a reflective collimator RC of $f=7$ mm focal length. The resulting low divergence Gaussian beam passes through a single order half-wave plate HWP, whose angular position allows to modify the orientation of the polarization direction of this beam with respect to that of a fixed linear polarizer LP$_0$. This allows an independent adjustment of the useful power of the light beam and its emission spectrum (central wavelength and line-width), while keeping a fixed orientation (TE or TM) to its polarization.

This polarized light beam is divided into two sub-beams by a non-polarizing cube splitter BS$_1$. The reflected beam is sent towards the sample to be characterized and the light flux retro-reflected by this sample is transmitted by the same cube splitter and forms the Signal channel (SIG). The beam transmitted through BS$_1$ forms the Reference channel (REF): this beam is retro-reflected by a hollow retro-reflector HRR which also laterally shifts it with respect to the incidence direction. These two channels are then superimposed through a second non-polarizing cube splitter BS$_2$ and the two complementary outputs of this coherent mixer are detected by the two photodiodes (PD$_1$ and PD$_2$) of a NIRVANA balanced receiver \cite{Hobbs_1997}. Four mirrors (from M$_1$ to M$_4$) and a total reflection prism (RAP, right angle prism) are used to adjust the position or orientation of the different beams inside the set-up.

From now on, we will assume that the sample is a flat silica window with an anti-reflective coating on one side. In a very general way, the currents delivered by each of the two photodiodes are thus described by the following equations \cite{Khan_2021,Khan_2022}:
\begin{equation}
I_1=I_{\text{dc,1}}+I_{\text{ac},1}\qquad I_2=I_{\text{dc,2}}-I_{\text{ac},2}
\end{equation}
with, for $j=1,2$:
\begin{equation}
I_{\text{dc},j}=\eta_aT_{\text{ref},j}\int\limits_0^{\infty}S(f)\mathcal{P}(f)\thinspace df+\eta_aT_{\text{sig},j}\int\limits_{0}^{\infty}S(f)\mathcal{P}(f)\left|r(f)\right|^2\thinspace df
\label{eq:Idc}
\end{equation}
and
\begin{equation}
I_{\text{ac},j}=2\eta_a\sqrt{T_{\text{ref},j}T_{\text{sig},j}}\thinspace\Re\left\{\int\limits_{0}^{\infty}S(f)\mathcal{P}(f)r(f)\thinspace e^{-ik_{\text{a}}\Delta L}\thinspace df\right\}
\label{eq:Iac}
\end{equation}
where $f$ is the frequency of the optical field, $S(f)$ is the spectral dependence of the photodiode responsivity, $\mathcal{P}(f)$ is the power spectral density of the light source, $r(f)$ is the coherent coefficient of reflection of the plane glass window \cite{Amra_2021}, $k_a$ is the wave vector in air, $\Delta L$ is the optical path difference between SIG and REF channels, $T_{\text{ref},j}$ (respectively $T_{\text{sig},j}$) is the transmittion coefficient of all the optical elements crossed by the reference (respectively signal) beam between the source and the photodiode $j$ ($j=1,2$), and $\eta_a$ is a factor that quantifies the geometrical overlap between the Gaussian profile of the light beam and the sensitive area of the photodiode, i.e.
\begin{equation}
\eta_a=\frac{\displaystyle\int\limits_0^{a}e^{-2r^2/w_d^2}r\thinspace dr}{\displaystyle\int\limits_0^{\infty}e^{-2r^2/w_d^2}r\thinspace dr}=1-e^{-2a^2/w_d^2}
\label{eq:Eta_a}
\end{equation}
where $a$ is the radius of the sensitive area of the photodiodes and $w_d$ is the modal radius of the Gaussian beam after a propagation distance $d$ from the exit pupil of the RC reflective collimator.

The RF output $V$ of the balanced receiver corresponds to the voltage resulting from the amplification of the difference between the currents of the two photodiodes whose dc components are balanced, i.e.
\begin{equation}
V=G(I_2-\alpha I_1)\quad\text{with}\quad I_{\text{dc},2}=\alpha I_{\text{dc},1}
\end{equation}
where $G$ is the transimpedance gain of the RF channel. LP$_1$ and LP$_2$ linear polarizers (in Fig. \ref{fig:BenchScheme}) are used to fine tune the balancing of the dc components. Consequently, we have
\begin{equation}
V=G\mathcal{T}\thinspace\Re\left\{\int\limits_{0}^{\infty}S(f)\mathcal{P}(f)r(f)\thinspace e^{-ik_a\Delta L}\thinspace df\right\}
\end{equation}
where $\mathcal{T}$ is a global transmission factor given by
\begin{equation}
\mathcal{T}=2\eta_a\left(\sqrt{T_{\text{ref},1}T_{\text{sig},1}}+\alpha\sqrt{T_{\text{ref},2}T_{\text{sig},2}}\right)
\end{equation}

\subsection{Data Processing}
\label{sec:DataProcessing}

\begin{figure}[htbp]
\centering\includegraphics[width=1.0\textwidth]{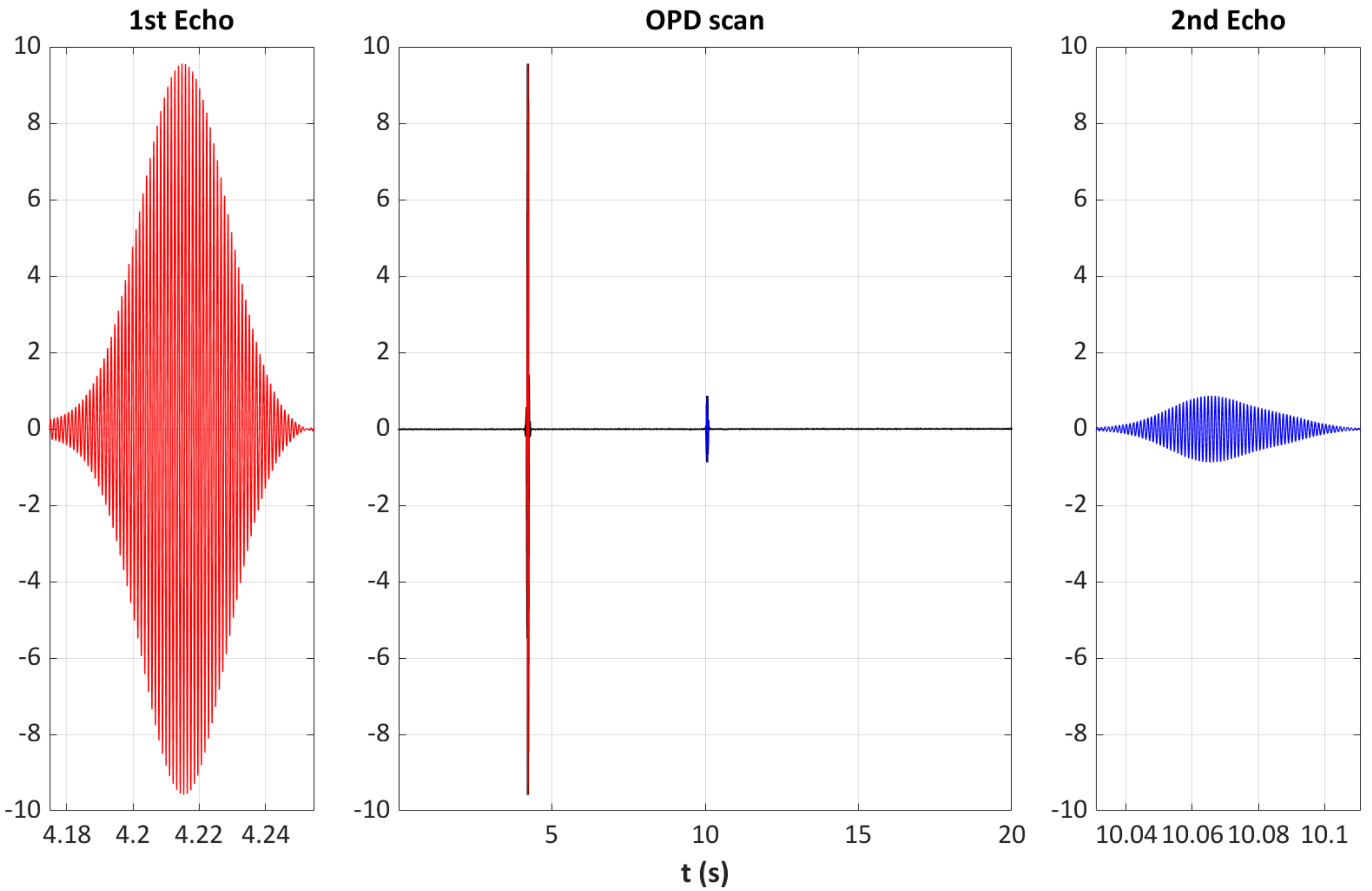}
\caption{Time dependence of the voltage $V$ recorded by the set-up when the front side of the silica window corresponds to the uncoated side (SLD driving current 190 mA, translation speed  0.5 mm/s, window thickness 2 mm) - Left, enlarged view of the first echo (uncoated face); center, full scan in optical path difference (OPD); right, enlarged view of the second echo (coated face).}
\label{fig:UNCARCScan}
\end{figure}

The voltage $V$ is digitized over 16 bits while the corner cube is translated at a constant speed $v$ along the $z$ axis. Figure \ref{fig:UNCARCScan} shows the time dependence of this digitized voltage when the front face of the sample corresponds to the uncoated face of the plane window (raw data).
The first echo (the resulting interference signal from a broadband light source) is obviously the highest amplitude signal, but the detection of the second echo is obtained with a very good signal-to-noise ratio (SNR), even when the anti-reflection coating is, as here, of very good quality (average reflection coefficient of about 300 ppm on the spectral bandwidth of the source).

\begin{figure}[htbp]
\centering\includegraphics[width=0.2\textwidth]{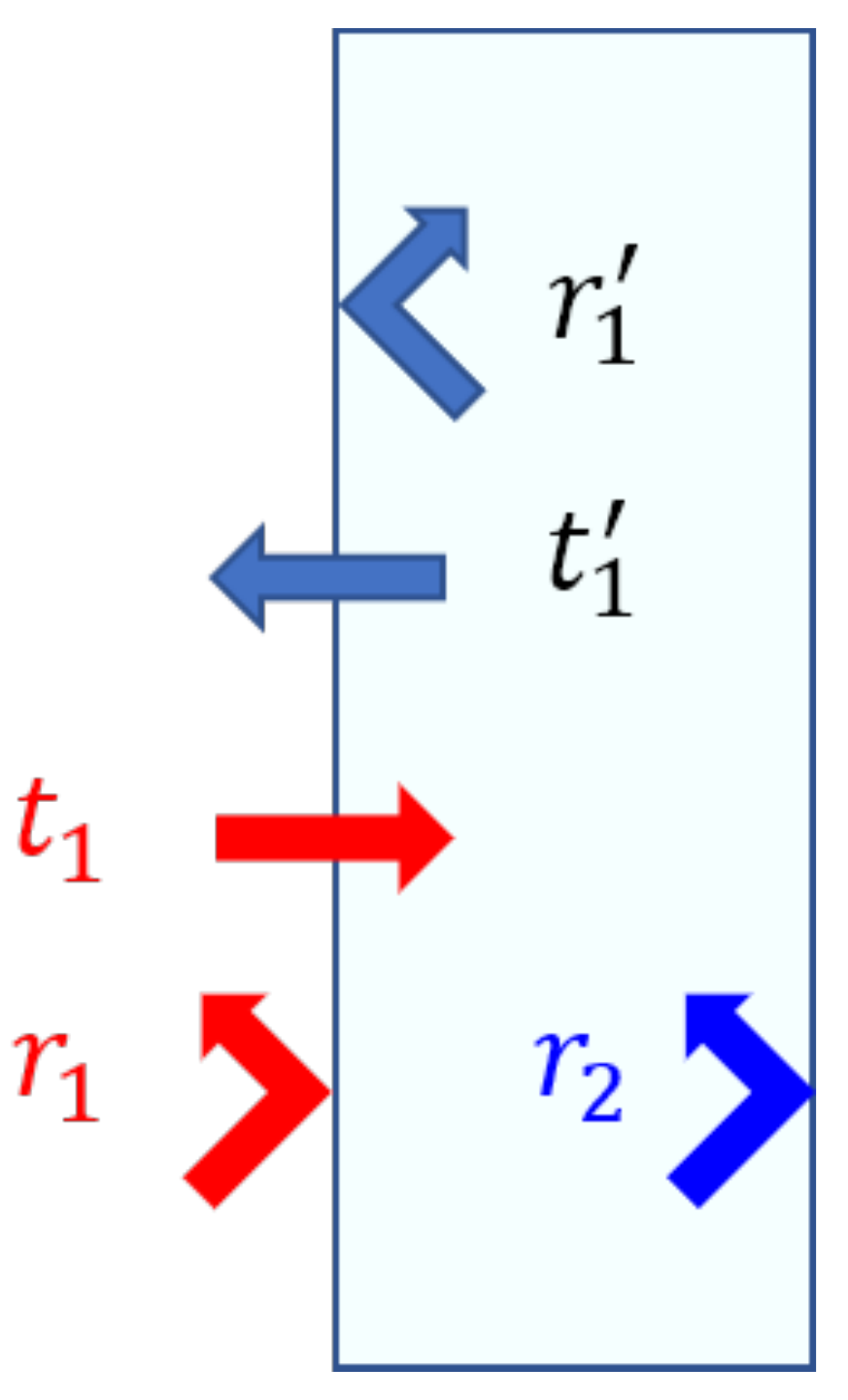}
\caption{Schematic view of the multiple reflections inside the plane-parallel window.}
\label{fig:Window_Scheme}
\end{figure}

To obtain the mathematical expression of the time dependence of the voltage $V$, we must take into account the uniform translational motion of the retroreflector used in the reference channel [$\Delta L=2vt$] as well as the frequency expression of the wave vector in air [$k_{\text{a}}\approx k_v=2\pi f/c$, where $k_v$ is the wave vector in vacuum], which leads to
\begin{equation}
V(t)=G\mathcal{T}\thinspace\Re\left\{\int\limits_{0}^{\infty}S(f)\mathcal{P}(f)r(f)\thinspace e^{-2i\pi\frac{2v}{c}ft}\thinspace df\right\}
\label{eq:V}
\end{equation}
Moreover, in the case of a window with plane and parallel faces, the reflection coefficient $r(f)$ can be put in the form of an infinite sum of elementary reflections, namely
\begin{equation}
r=r_1+t_1r_2t'_1\thinspace e^{2ik_vd_sn_s}+t_1r_2r'_1r_2t'_1\thinspace e^{4ik_vd_sn_s}+...
\label{eq:rf}
\end{equation}
where $d_s$ is the thickness of the window, $n_s$ is its refractive index, and the coefficients $r$, $r'$, $t$, and $t'$ are as shown in Fig. \ref{fig:Window_Scheme}. The frequency dependence of these quantities has been omitted here for the sake of simplicity.

Combining (\ref{eq:V}) and (\ref{eq:rf}), we get
\begin{equation}
V(t)=\sum\limits_{m=1}^{\infty}V_m(t)=\sum\limits_{m=1}^{\infty}G\mathcal{T}\thinspace\Re\left\{\int\limits_{0}^{\infty}\mathcal{B}_m(f)\thinspace e^{-2i\pi\frac{2v}{c}ft}\thinspace df\right\}
\label{eq:V_Vm}
\end{equation}
where
\begin{equation}
\mathcal{B}_m(f)=S(f)\mathcal{P}(f)\rho_m(f)\quad\text{with}\quad\left\{
\begin{aligned}
&\rho_1=r_1\\
&\rho_2=t_1r_2t'_1\thinspace e^{2i\pi\frac{2d_sn_s}{c}f}\\
&\rho_3=t_1r_2r'_1r_2t'_1\thinspace e^{2i\pi\frac{4d_sn_s}{c}f}\\
&...
\end{aligned}
\right.
\label{eq:Jm}
\end{equation}

The quantities $S(f)$ and $\mathcal{P}(f)$ are real functions with bounded support in $\mathbb{R}^+$ and the spectral profile of their product is very close to a Gaussian centered at $f_0=c/\lambda_0$ and whose full width at half maximum is $\Delta f$ \cite{Khan_2021}. This boundedness allows us to replace the lower limit of integration of equation (\ref{eq:V_Vm}), i.e. 0, by $-\infty$. Each function $V_m(t)$ is therefore proportional to the real part of the Fourier transform of a Gaussian, whose shape is both strongly attenuated and slightly modulated (a single oscillation within the frequency support) by the reflection coefficient $r_2(f)$. Consequently, the total width $\Delta t$ of the support of this function is defined in order of magnitude by 
\begin{equation}
\Delta\left(\frac{2v}{c}t\right)\sim 4\frac{4\pi}{\Delta f}\quad\Rightarrow\quad\Delta t\sim\frac{8\pi\lambda_0^2}{v\thinspace\Delta\lambda}
\end{equation}
The spectral width $\Delta\lambda$ of the superluminescent diode is an increasing function of the driving current and varies between 27 nm and 82 nm, full width at half maximum (FWHM), while its central wavelength $\lambda_0$ varies correspondingly between 1076 nm and 1042 nm. Therefore, the width of the support of the functions $V_m(t)$ is, in the worst case, on the order of $1/v$ seconds for $v$ in mm/s.

Besides, the time interval $\Delta T$ separating two consecutive echoes is defined by \cite{Khan_2021}
\begin{equation}
\Delta T=\frac{n_g(\lambda_0)d_s}{v}
\end{equation}
where $n_g$ is the group index of the window glass. In order to ensure no overlap of the functions $V_m(t)$, the following condition must be satisfied
\begin{equation}
\Delta T>\Delta t\quad\Rightarrow\quad n_g(\lambda_0)d_s>1\text{ mm}
\end{equation}
or, for a silica window: $d_s>0.7$ mm. The samples we use have thicknesses of 2 mm, so this non-overlapping condition is largely respected.

The data processing that we implement consists of 
\begin{enumerate}
\item isolating in the signal $V(t)$ temporal windows of width $\Delta T$ centered on each of the echoes
\begin{equation}
W_m(t)=\text{Rect}\left[\frac{t-t_m}{\Delta T}\right]V(t)\quad\text{where}\quad t_m=(m-1)\Delta T
\end{equation}
\item using the non-overlapping condition to replace $W_m(t)$ by $V_m(t)$
\item calculating numerically the discrete Fourier transform (DFT) of the windowed signals $V_m(t)$, i.e.
\begin{equation}
\mathcal{S}_m(F_l)=\sum\limits_{k=-N/2}^{k=N/2-1}V_m(t_k)\thinspace e^{-2i\pi F_lt_k}\thinspace dt\quad\text{for }l=-N/2,-N/2+1,...,N/2-1
\label{eq:TFVm}
\end{equation}
where
\begin{equation}
dt=\frac{1}{F_s}=\frac{\Delta T}{N-1}\quad\text{;}\quad t_k=t_m+k.dt\quad\text{;}\quad F_l=l.dF\quad\text{;}\quad dF=\frac{1}{(N-1)dt}=\frac{1}{\Delta T}
\end{equation}
This discrete Fourier transform is associated with a continuous Fourier transform defined by
\begin{equation}
\widetilde{S}_m(F)=\int\limits_{-\infty}^{+\infty}V_m(t)\thinspace e^{-2i\pi Ft}\thinspace dt = G\mathcal{T}\int\limits_{-\infty}^{+\infty}\Re\left\{\int\limits_{-\infty}^{+\infty}\mathcal{B}_m(f)\thinspace e^{-2i\pi\frac{2v}{c}ft}\thinspace df\right\}\thinspace e^{-2i\pi Ft}\thinspace dt
\end{equation}
which can be easily calculated by transforming the real part into a half-sum of conjugated complex quantities, which leads to
\begin{equation}
\widetilde{V}_m(F)=\frac{1}{2}G\mathcal{T}\left\{\frac{c}{2v}\mathcal{B}_m\left(-\frac{c}{2v}F\right)+\frac{c}{2v}\mathcal{B}_m^*\left(\frac{c}{2v}F\right)\right\}
\label{eq:ContinuousFT}
\end{equation}
\item using the latter result to identify the DFT terms $\mathcal{S}_m(F_l)$ with the continuous Fourier transform $\widetilde{V}_m(F)$ sampled at $F=F_l$, or
\begin{equation}
\mathcal{S}_m(F_l)=G\mathcal{T}\frac{c}{4v}\mathcal{B}_m^*\left(\frac{c}{2v}F_l\right)=G\mathcal{T}\frac{c}{4v}S(f_l)\mathcal{P}(f_l)\rho_m^*(f_l)\quad\text{where}\quad f_l=\frac{c}{2v}F_l
\end{equation}
\item removing the unknown terms by taking the ratio between two DFT samples, one of which is chosen as calibration term
\begin{equation}
\frac{\mathcal{S}_m^*(F_l)}{\mathcal{S}_c^*(F_l)}=\frac{\rho_m(f_l)}{\rho_c(f_l)}
\label{eq:Echo_Calibration}
\end{equation}
If we assume, for example, that the front face of the window corresponds to the uncoated side, we will choose the first echo as calibration echo ($c=1$), which will allow us to write
\begin{equation}
\frac{|\mathcal{S}_2(F_l)|^2}{|\mathcal{S}_1(F_l)|^2}=\frac{|\rho_2(f_l)|^2}{|\rho_1(f_l)|^2}=\frac{|t_1(f_l)t'_1(f_l)r_2(f_l)|^2}{|r_1(f_l)|^2}=\frac{T_1^2(f_l)R_2(f_l)}{R_1(f_l)}
\label{eq:FFTSquareModulusRatio}
\end{equation}
or
\begin{equation}
R_{\text{coat}}(f_l)=\frac{R_s(f_l)}{[1-R_s(f_l)]^2}\frac{|\mathcal{S}_2(F_l)|^2}{|\mathcal{S}_1(F_l)|^2}\quad\text{where}\quad R_s(f_l)=\left[\frac{n_s(f_l)-1}{n_s(f_l)+1}\right]^2
\label{eq:Rcoat}
\end{equation}
The spectral dependence of the refractive index of the substrate being perfectly known, the ratio of the power spectral densities of the Fourier transforms of the first two echoes allows us to determine the spectral dependence of the reflection coefficient of the coated side. If we now reverse the orientation of the window, the same approach leads to
\begin{equation}
\frac{R_{\text{coat}}(f_l)}{[1-R_{\text{coat}}(f_l)]^2}=R_s(f_l)\frac{|\mathcal{S}_1(F_l)|^2}{|\mathcal{S}_2(F_l)|^2}
\end{equation}
which is a bite more complicated to process from a numerical point of view than the result obtained with the first orientation. But, with this second orientation, we can also write
\begin{equation}
\text{arg}[\rho_1(f_l)]=\text{arg}[r_{\text{coat}}(f_l)]=-\text{arg}[\mathcal{S}_1(F_l)]
\label{eq:Phase}
\end{equation}
and thus determine, in a very simple way, the spectral dependence of the phase shift on the anti-reflection coating. Note that all these measurements are performed at very low frequencies ($F$ is about 2 kHz for a translation speed $v$ of 1 mm/s), while the results obtained are at optical frequencies ($f$ about 300 THz). This frequency down-conversion is one of the key advantages of Fourier transform spectrometry \cite{Fellgett_1958,Connes_1966}.
\end{enumerate}

As discussed in the introduction, the processing method that we propose allows us to determine the spectral dependence of the reflection coefficient of the coated interface, in amplitude and phase. However, it remains to be answered the achieved precision and spectral resolution using this method. And in order to address the first point, we have a dedicated section (\ref{sec:DetectionNoise}) due to the involved complexity, while the second quantity can be quickly estimated from the description of the processing method that we have just described. Indeed, the DFT samples $\mathcal{S}_m(F_l)$ introduced in (\ref{eq:TFVm}) can be expressed in terms of the continuous Fourier transform $\widetilde{V}_m(F)$ as \cite{Thurman_2007}
\begin{multline}
\mathcal{S}_m(F_l)=\frac{1}{dt\sqrt{N}}\int\limits_{-\infty}^{+\infty}\left\{\widetilde{V}_m(F)\star\left[\frac{1}{dF}\text{sinc}\left(\frac{F}{dF}\right)\right]\right.\\
\left.\star\left[\frac{1}{NdF}\text{comb}\left(\frac{F}{NdF}\right)\right]\right\}\delta(F-l.dF)\thinspace dF
\label{eq:LienDFTFT}
\end{multline}
where the $\star$ symbol represents a convolution operation, sinc is the sine cardinal function [$\text{sinc}(x)=\sin(\pi x)/(\pi x)$] and comb is the Dirac comb function. The presence of a convolution by a cardinal sine in equation (\ref{eq:LienDFTFT}) shows that the spectral resolution of this method is defined by the frequency pitch $dF$, that is
\begin{equation}
dF=\frac{1}{\Delta T}\quad\Rightarrow\quad df=\frac{c}{2v\Delta T}\quad\Rightarrow\quad d\lambda=\frac{\lambda_0^2}{2n_gd_s}
\end{equation}
or 0.2 nm for a 2mm-thick silica window.

\subsection{Sensitivity to detection noise and alignment bias}
\label{sec:Sensitivity}

\subsubsection{Detection noise}
\label{sec:DetectionNoise}

The sources of noise which could affect the measurement of a reflection coefficient $|\rho_m|^2$ associated with the echo $m$ are essentially the quantum noise associated with the DC component of the current delivered by each photodiode and the residual effect of the intensity noise of the superluminescent diode. This last term can result from the imperfect rejection of the common modes of disturbance which reflects the CMRR (Common Mode Rejection Ratio) of the balanced receiver.

The variance of the shot noise affecting the voltage $V$ provided by the balanced receiver is proportional to the sum of the quantum noise affecting each of the two photodiodes (the two noises are indeed independent), i.e.
\begin{equation}
\sigma_I^2=2e(I_1+I_2)\thinspace B\quad\Rightarrow\quad\sigma_V^2=G^2\sigma_I^2\sim 2G^2e(I_{\text{dc},1}+I_{\text{dc},2})\thinspace B=2G^2(1+\alpha)eI_{\text{dc},1}\thinspace B
\label{eq:BruitShot}
\end{equation}
where $B$ is the detection bandwidth and $e$ the elementary charge. We must also consider the contribution related to the dark current $I_{\text{dark}}$ of each of these photodiodes, or
\begin{equation}
\sigma_I^2=4eI_{\text{dark}}\thinspace B\quad\Rightarrow\quad \sigma_V^2=4G^2eI_{\text{dark}}\thinspace B\sim2G^2S^2\text{NEP}^2\thinspace B
\label{eq:BruitIntrinseque}
\end{equation}
where $S$ is the responsivity of the photodiode ($S\sim 0.8$ A/W) and NEP its noise equivalent power (3 pW/$\sqrt{\text{Hz}}$). If we only consider noise of quantum origin, the variance of $V$ is thus defined by
\begin{equation}
\sigma_V^2=2G^2(1+\alpha)eI_{\text{dc},1}\thinspace B+2G^2S^2\text{NEP}^2\thinspace B
\label{eq:sigma2Vq}
\end{equation}
while the corresponding signal to noise ratio is written
\begin{equation}
\text{SNR}_q=\frac{V^2}{\sigma_V^2}=\frac{G^2(I_{\text{ac,2}}+\alpha I_{\text{ac},1})^2}{2G^2(1+\alpha)eI_{\text{dc},1}\thinspace B+2G^2S^2\text{NEP}^2\thinspace B}
\label{eq:SNRq}
\end{equation}
We will now assume that the set-up is spontaneously balanced ($\alpha=1$) and take into account that the power detected by the receiver must not exceed a maximum value $P_{\text{max}}$, due either to the saturation of the two photodiodes or the digitizing range of the voltage $V$.

Initially, assume that this maximum value is defined by the absence of saturation of photodiodes. Consequently
\begin{equation}
I_{\text{dc},1}=SP_{\text{sat}}
\label{eq:P_sat}
\end{equation}
In the case of a window, anti-reflection coated on one side, multiple reflections are dominated by the one that occurs on the uncoated side in simple bounce. Therefore, equation (\ref{eq:Idc}) becomes
\begin{equation}
I_{\text{dc},1}=\eta_a\left\{T_{\text{ref},1}+T_{\text{sig},1}R_{\text{uncoat}}\right\}SP
\label{eq:Idc1}
\end{equation}
where $P$ is the total power emitted by the source. The main difference between the signal and reference channels is the presence of an additional reflection on the BS$_1$ cube splitter in the case of the signal channel. Therefore, to a first approximation
\begin{equation}
T_{\text{sig,1}}\sim\frac{T_{\text{ref,1}}}{2}
\label{eq:Tsig1}
\end{equation}
Combining (\ref{eq:P_sat}), (\ref{eq:Idc1}) and (\ref{eq:Tsig1}), we get
\begin{equation}
P=\frac{P_{\text{sat}}}{\eta_aT_{\text{ref},1}(1+R_{\text{uncoat}}/2)}\sim\frac{P_{\text{sat}}}{\eta_aT_{\text{ref},1}}
\label{eq:P_psat}
\end{equation}
To conclude, we need to know:
\begin{itemize}
\item[\textbullet] the value of the geometric overlap factor $\eta_a$; the modal radius $w_d$ of the Gaussian beam after a propagation over a distance $d$ is given by :
\begin{equation}
w_d=w_f\sqrt{1+\left(\frac{\lambda d}{\pi w_f^2}\right)^2}\quad\text{with}\quad w_f=\frac{f\lambda}{\pi w_0}
\label{eq:wd}
\end{equation} 
In our set-up, $d=1275$ mm, $w_f=0.67$ mm, and $w_d=0.92$ mm, which leads to a geometric overlap factor $\eta_a$ of about 0.44.
\item[\textbullet] the transmission of the reference channel; as can be seen in Fig. \ref{fig:BenchScheme}, we essentially have to consider two crossings of a splitter cube ($\times 0.5$ each), two crossings of a polarizer ($\times 0.87$ each) and four reflections on a silver coating ($\times 0.95$ each). Therefore
\begin{equation}
T_{\text{ref,1}}=(0.5)^2\times(0.87)^2\times(0.95)^4\sim 0.15
\end{equation}
\end{itemize}
The saturation power of the Nirvana receiver is 0.5 mW. So, to reach saturation, the total power $P$ delivered by the superluminescent diode must be equal to 7.5 mW, which corresponds to a driving current of about 180 mA (for a maximum value of 1000 mA).

Now assume that the maximum power is defined by the digitizing range of the voltage $V$, which is 10 volts. At the top of echo $m$, it follows that
\begin{equation}
V=G(I_{\text{ac},1}+\alpha I_{\text{ac,2}})=G\mathcal{T}SP|\rho_m|\leqslant 10
\end{equation}
where
\begin{equation}
\mathcal{T}\approx 4\eta_a\sqrt{T_{\text{ref},1}T_{\text{sig},1}}\approx 2\sqrt{2}\eta_aT_{\text{ref},1}
\label{eq:MathcalT}
\end{equation}
The highest amplitude echo is obviously the one corresponding to the uncoated face, and therefore
\begin{equation}
P\leqslant\frac{10}{2\sqrt{2}\eta_aT_{\text{ref},1}GS\sqrt{R_s}}\sim 3.6\text{ mW}
\label{eq:PmaxADC}
\end{equation}

This last condition is the most restrictive one, and it thus defines the maximum power $P_{\text{max}}$ on the photodiodes, namely
\begin{equation}
P_{\text{max}}\approx\eta_aT_{\text{ref},1}P\sim 250\text{ $\mu$W}
\end{equation}
Accordingly
\begin{equation}
I_{\text{dc},1}=SP_{\text{max}}\quad\text{;}\quad I_{\text{ac},1}=\frac{S}{\sqrt{2}}|\rho_m|P_{\text{max}}
\label{eq:Idc1Iac1}
\end{equation}
By combining the equations (\ref{eq:SNRq}) and (\ref{eq:Idc1Iac1}), we obtain the following final expression for the signal-to-noise ratio
\begin{equation}
\text{SNR}_q=\frac{S^2|\rho_m|^2P_{\text{max}}^2}{(2eSP_{\text{max}}+S^2\text{NEP}^2)\thinspace B}
\end{equation}

The smallest value of the reflection coefficient $|\rho_m|^2$ that we are able to measure with a signal to noise ratio of 10 is therefore defined by
\begin{equation}
|\rho_m|_q^2=10\frac{2eSP_{\text{max}}+S^2\text{NEP}^2}{S^2P_{\text{max}}^2}B\approx10\frac{2e}{SP_{\text{max}}}B\sim 1.6\times 10^{-14}B
\end{equation}
or $2\times 10^{-9}$ if the balanced receiver is used at its maximum bandwidth ($B=125$ kHz).

Using a similar approach, we can estimate the minimum value of the reflection coefficient that can be detected in the presence of a residual impact of source intensity noise. If the balanced receiver were operating perfectly, the source intensity noise would not affect the voltage $V$. But the rejection of these correlated noise sources is not perfect, which is quantified by the measure of CMRR in balanced photodetection. Therefore
\begin{equation}
\sigma_V^2=G^2\sigma_I^2=G^2S^2\sigma_P^2=G^2\times 10^{(\text{RIN}-\text{CMRR})/10}S^2P_{\text{max}}^2B
\label{eq:sigma2RIN}
\end{equation}
where RIN is the relative intensity noise of the source ($-105$ dB/Hz for the SLD) and the maximum attainable CMRR of NIRVANA receiver is 50 dB.

The expression of the signal $V$ is identical to that established in the shot noise study, i.e.
\begin{equation}
V=G(I_{\text{ac},2}+\alpha I_{\text{ac},1})\approx2GI_{\text{ac},1}=2\sqrt{2}G\eta_aT_{\text{ref},1}S|\rho_m|P=2\sqrt{2}GS|\rho_m|P_{\text{max}}
\end{equation}

Therefore, the signal-to-noise ratio is expressed as
\begin{equation}
\text{SNR}_{\text{RIN}}=\frac{V^2}{\sigma_V^2}=\frac{8|\rho_m|^2}{10^{(\text{RIN}-\text{CMRR})/10}B}
\end{equation}
and the smallest value of the reflection coefficient $|\rho_m|^2$ that we are able to measure with a signal to noise ratio of 10 this time is defined by
\begin{equation}
|\rho_m|_{\text{RIN}}^2\sim 10^{(\text{RIN}-\text{CMRR})/10}B
\end{equation}
or $4\times 10^{-11}$ if the balanced receiver is used at its maximum bandwidth ($B=125$ kHz). This result is important because it shows that the resolution of the measurement will remain limited by quantum noise for any CMRR value between 35 dB and 50 dB. We will now assume that this condition is satisfied in our theoretical estimation.

However, all the results we have just presented are related to the direct use of the measurement signal $V(t)$, and not to its discrete Fourier transforms $\mathcal{S}_m(F_l)$, as defined by equation (\ref{eq:TFVm}). It is therefore necessary to take into account this key step of processing in the estimation of the performance of our measurement method.

In the expression of the discrete Fourier transform (\ref{eq:TFVm}), let us make the changes of variable
\begin{equation}
\bar{t}_m=t_m-\frac{N}{2}dt\quad\text{,}\quad p=k+\frac{N}{2}\quad\text{and}\quad t_p=\bar{t}_m+p.dt
\end{equation}
which leads to
\begin{equation}
\mathcal{S}_{m,l}=e^{-2i\pi l\bar{t}_m dF}\sum\limits_{p=0}^{p=N-1}V_{m,p}\thinspace e^{-2i\pi lpdFdt}\thinspace dt\quad\text{where}\quad dFdt=\frac{1}{(N-1)}
\end{equation}
This equation can be put in the following matrix form
\begin{equation}
\vec{\mathcal{S}}_m=\textbf{A}.\vec{V}_m\quad\text{or}\quad \mathcal{S}_{m,l}=\sum\limits_{p=0}^{p=N-1}a_{lp}^{}V_{m,p}\quad\text{where}\quad a_{lp}^{}=e^{-2i\pi l\bar{t}_m dF}\thinspace e^{-2i\pi\frac{lp}{N-1}}\thinspace dt
\end{equation}
In the presence of shot noise on the measurement of the voltage $V(t)$, i.e. on the components of the vector $\vec{V}_m$, this matrix equation becomes
\begin{equation}
\vec{\mathcal{S}}_m+\vec{n}'_m=\textbf{A}.(\vec{V}_m+\vec{n}_m)
\end{equation}
where $\vec{n}_m$ and $\vec{n}'_m$ are the noise vectors affecting respectively the measurement of the vectors $\vec{V}_m$ and $\vec{\mathcal{S}}_m$. We then introduce the covariance matrices of these noise vectors, defined by \cite{Matallah_2011}
\begin{equation}
\textbf{$\Gamma$}^{n}_m=\langle\vec{n}_m.^t\vec{n}_m^{*}\rangle\quad\text{;}\quad\textbf{$\Gamma$}^{n'}_m=\langle\vec{n'}_m.^t\vec{n'}_m^{*}\rangle
\end{equation}
where the bracketing denotes an ensemble averaging, $^t$ the transpose and $^*$ the complex conjugation. Using the definition of the vector $\vec{n}'$, it becomes
\begin{equation}
\textbf{$\Gamma$}^{n'}_m=\langle\textbf{A}.\vec{n}_m.^t\vec{n}_m^*.^t\textbf{A}^*\rangle=\langle\textbf{A}.\textbf{$\Gamma$}^{n}_m.^t\textbf{A}^*\rangle
\end{equation}
The noise affecting the measurement of the components of the vector $\vec{n}$ is a Gaussian white noise, of mean zero and variance $\sigma^2_V$ defined by [see (\ref{eq:sigma2Vq})]
\begin{equation}
\sigma_V^2=4G^2eSP_{\text{max}}B+2G^2S^2\text{NEP}^2B=2G^2\left\{2eSP_{\text{max}}+S^2\text{NEP}^2\right\}B
\label{eq:sigma2q}
\end{equation}
The covariance matrix $\textbf{$\Gamma$}^{n}_m$ is diagonal \cite{Matallah_2011}. Moreover, this variance depends neither on the component of the vector $\vec{V}_m$, nor on the order $m$ of the echo. Therefore
\begin{equation}
\textbf{$\Gamma$}^{n}_m=\sigma_V^2\textbf{$\mathbb{I}$}
\end{equation}
where $\mathbb{I}$ is the identity matrix. We deduce the expression of the covariance matrix $\textbf{$\Gamma$}^{n'}_m$, that is
\begin{equation}
\textbf{$\Gamma$}^{n'}_m=\sigma_V^2\langle\textbf{A}.^t\textbf{A}^*\rangle
\end{equation}
The elements of the covariance matrix $\textbf{$\Gamma$}^{n'}_m$ are therefore defined by
\begin{equation}
[\textbf{$\Gamma$}^{n'}_m]_{lp}^{}=e^{-2i\pi (l-p)\bar{t}_m dF}\sum\limits_{q=0}^{N-1} \thinspace e^{-2i\pi\frac{q(l-p)}{N-1}}\thinspace\sigma_V^2 (dt)^2
\end{equation}
or, for the diagonal elements, the only ones necessary to estimate the noise affecting the measurement of $\mathcal{S}_{m,l}$ \cite{Matallah_2011,Ferrec_2008}
\begin{equation}
[\textbf{$\Gamma$}^{n'}_m]_{ll}^{}=N\sigma_V^2 (dt)^2
\end{equation}

Finally, the signal to noise ratio of our measurement is
\begin{equation}
\text{SNR}_{\mathcal{S}_{m,l}}=\frac{|\mathcal{S}_{m}(F_l)|^2}{[\textbf{$\Gamma$}^{n'}_m]_{ll}^{}}=\frac{|G\mathcal{T}\frac{c}{4v}\mathcal{B}_m^*\left(\frac{c}{2v}F_l\right)|^2}{N\sigma_V^2 (dt)^2}=\left[G\mathcal{T}\frac{c}{4v}\right]^2\frac{|S(f_l)\mathcal{P}(f_l)\rho_m^*(f_l)|^2}{N\sigma_V^2 (dt)^2}
\end{equation}
where
\begin{equation}
f_l=\frac{c}{2v}F_l\quad F_l>0\quad\text{;}\quad P_{\text{max}}=\eta_aT_{\text{ref},1}\int\limits_0^{\infty}\mathcal{P}(f)\thinspace df=\int\limits_0^{\infty}\mathcal{P}_{\text{max}}(f)\thinspace df
\end{equation}
Using the expression for the variance of the noise affecting the interferogram measurement defined by (\ref{eq:sigma2q}), we get
\begin{equation}
\text{SNR}_{\mathcal{S}_{m,l}}=\left[\mathcal{T}\frac{c}{4v}\right]^2\frac{S^2\mathcal{P}^2(f_l)}{2(2eSP_{\text{max}}+S^2\text{NEP}^2)B}\frac{|\rho_m(f_l)|^2}{N(dt)^2}
\end{equation}
In addition
\begin{equation}
\frac{1}{N(dt)^2}=\frac{1}{Ndt\times dt}=\frac{F_s}{\Delta T}
\label{eq:dt}
\end{equation}
Using the equations (\ref{eq:MathcalT}) and (\ref{eq:dt}), the signal-to-noise ratio can be written as follows
\begin{equation}
\text{SNR}_{\mathcal{S}_{m,l}}=\left[\frac{c}{2v}\right]^2\frac{S^2\mathcal{P}_{\text{max}}^2(f_l)}{(2eSP_{\text{max}}+S^2\text{NEP}^2)B}\frac{F_s}{\Delta T}\thinspace|\rho_m(f_l)|^2
\end{equation}
The smallest detectable reflection coefficient with a signal-to-noise ratio of at least 10 will furthermore be defined by
\begin{equation}
|\rho_m(f_l)|_q^2=10\left[\frac{2v}{c}\right]^2\frac{(2eSP_{\text{max}}+S^2\text{NEP}^2)B}{S^2\mathcal{P}_{\text{max}}^2(f_l)}\frac{\Delta T}{F_s}
\label{eq:rhomin}
\end{equation}
and thus depends on the optical frequency $f_l$ through the saturation power spectral density $\mathcal{P}_{\text{max}}(f_l)$.

Let us suppose that this power spectral density is a rectangle function of width $\Delta f$. Then
\begin{equation}
P_{\text{max}}=\mathcal{P}_{\text{max}}.\Delta f=\mathcal{P}_{\text{max}}\frac{c\Delta\lambda}{\lambda_0^2}\quad\Rightarrow\quad c\mathcal{P}_{\text{max}}=\frac{\lambda_0^2}{\Delta\lambda}P_{\text{max}}
\end{equation}
The equation (\ref{eq:rhomin}) then becomes
\begin{equation}
|\rho_m(f_l)|_q^2=10\left(\frac{\Delta\lambda}{\lambda_0}\right)^2\frac{(2eSP_{\text{max}}+S^2\text{NEP}^2)B}{S^2 P_{\text{max}}^2}\frac{F_0^2}{F_s}\Delta T\approx10\left(\frac{\Delta\lambda}{\lambda_0}\right)^2\frac{2eB}{SP_{\text{max}}}\frac{F_0^2}{F_s}\Delta T
\label{eq:rhomin_2}
\end{equation}
Using the numerical values listed below
\begin{center}
$\lambda_0$ = 1068 nm ; $\Delta\lambda$ = 33 nm\\
$e=1.6\times 10^{-19}$ C ; $S=0.8$ A/W ; $P_{\text{max}}=0.25$ mW ; NEP = 3 pW/$\sqrt{\text{Hz}}$\\
$B=125$ kHz ; $v=1$ mm/s ; $F_0=1.87$ kHz\\
$F_s=100$ kHz ; $\Delta T=2.92$ s
\end{center}
we find that the smallest measurable reflection coefficient with a signal-to-noise ratio of 10 is on the order of $2\times 10^{-10}$ over a spectral band of 30 nm, with a spectral resolution of 0.2 nm.

Note that if we reduce the duration $\Delta T$ of the processing window by a factor of 10, this detection floor is lowered by the same ratio, and thus reaches $2\times 10^{-11}$. The penalty to pay is the degradation of the spectral resolution by the same factor, i.e. 2 nm.

\subsubsection{Alignment bias}
\label{sec:AlignementBias}

So far, we have implicitly assumed that the window has its two sides perfectly parallel and perpendicular to the direction of the incident beam. However, in practice, these assumptions are most likely not verified, and we therefore need to analyze the possible consequences of small alignment bias (typically about 10 arc seconds).

\begin{figure} [htbp]
\centering\includegraphics[width=0.5\textwidth]{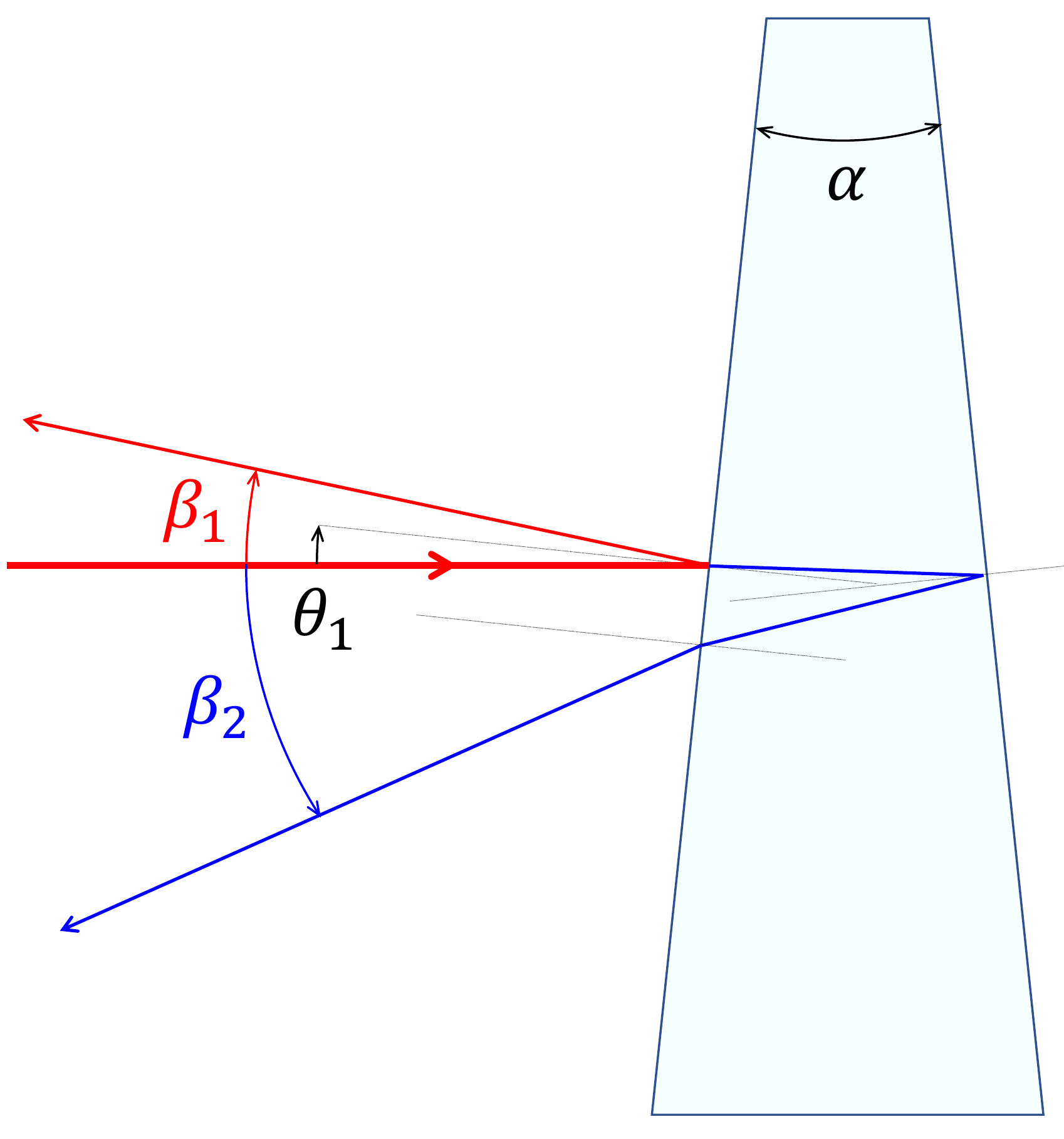}
\caption{Influence of the wedge and misalignment on the direction of the beams reflected from the two window faces}
\label{fig:Alignment_Bias}
\end{figure}

The most general situation is schematically represented in Fig. \ref{fig:Alignment_Bias}, where $\beta_1$ and $\beta_2$ are the angles between the incident beam and the beams respectively reflected from the front and rear faces of the window. 
All angles here are very small, so we can replace their sine by the value of the angle in radians. The angles are counted in a positive clockwise direction. If the wedge angle is included in the plane of incidence (which is the worst case), then it is easy to show that
\begin{equation}
\beta_1=2\theta_1\quad\text{;}\quad\beta_2=\beta_1-2n_s\alpha
\end{equation}

When the wedge angle $\alpha$ of the window is zero, the two angles $\beta_1$ and $\beta_2$ are equal, and the truncation effects induced on the spatial distribution of illumination of the two beams by the small size of the photodiodes (1 mm) are thus identical, and taken into account by the calibration procedure described in the section \ref{sec:DataProcessing}. Note that the beam reflected by the rear face has a lateral shift compared to that reflected by the front face, but it is extremely small (the lever arm is indeed the thickness of the window, or 2 mm), and can therefore be neglected.

On the other hand, when the wedge angle of the window is not zero, the beam reflected from the rear face has an angular bias $\delta\beta=-2n_s\alpha$ with respect to the beam reflected by the front face. This angular bias $\delta\beta$ causes the appearance of a gap $\delta x$ between the centroids of the two Gaussian beams during their detection by the photodiodes. This offset is defined by $\delta x=d.\delta\beta$, where $d$ is the propagation distance between the emission of the beam and its reception [$d\sim 1275$ mm, see equation (\ref{eq:wd})]. If the window has, for example, a wedge angle $\alpha$ of 2 arc seconds, the relative displacement $\delta x$ will be on the order of 35 $\mu$m, which cannot be neglected a priori. Moreover, since we are interested in measuring the spectral dependence of the reflection coefficient, our modeling of the consequences of this lateral shift must take into account all possible spectral dependencies.

In the presence of a misalignment $\theta_1$ and a wedge angle $\alpha$, the overlap factor $\eta_a$ will depend on both the order $m$ of the echo and the wavelength $\lambda$, and is written
\begin{equation}
\eta_{a,m}(\lambda)=\frac{2}{\pi w_d^2(\lambda)}\displaystyle\iint\limits_{\mathcal{C}}e^{\displaystyle -2\frac{[x-\beta_m d]^2+y^2}{w_d^2(\lambda)}}dxdy
\end{equation}
where $\mathcal{C}$ is a disk with center (0,0) and radius $a$, and $w_d(\lambda)$ is defined by equation (\ref{eq:wd}), in which the wavelength dependence of the modal radius $w_0$ is now considered, namely \cite{Marcuse_1978}
\begin{equation}
w_0(\lambda)=r_0\left(0.65+\frac{1.619}{[V(\lambda)]^{3/2}}+\frac{2.879}{[V(\lambda)]^6}\right)
\label{eq:MFD}
\end{equation}
where $r_0$ is the core radius, and $V$ the normalized frequency
\begin{equation}
V(\lambda)=\frac{2\pi r_0\times\text{NA}}{\lambda}=2.405\frac{\lambda_c}{\lambda}
\label{eq:NormalizedFrequency}
\end{equation}
NA being the numerical aperture of the fiber, and $\lambda_c$ its cutoff wavelength, below which it is no longer single mode. In our case (PM980-XP), $2r_0=5.5$ $\mu$m, $\lambda_c=870$ nm, and NA = 0.12.

Under these conditions, equation (\ref{eq:Echo_Calibration}) becomes
\begin{equation}
\frac{\mathcal{S}_m^*(F_l)}{\mathcal{S}_c^*(F_l)}=\frac{\eta_{a,m}(f_l)\rho_m(f_l)}{\eta_{a,c}(f_l)\rho_c(f_l)}\quad\text{where}\quad\lambda_l=\frac{c}{f_l}
\end{equation}
or, in the case of the determination of the spectral dependence of the reflection coefficient of the coated face
\begin{equation}
R_2^{\text{exp}}(\lambda_l)=\left[\frac{\eta_{a,1}(\lambda_l)}{\eta_{a,2}(\lambda_l)}\right]^2R_2(\lambda_l)=\mathcal{K}(\theta_1,\lambda_l;\alpha)R_2(\lambda_l)
\end{equation}
The error induced on the measurement is therefore multiplicative and the corrective factor $\mathcal{K}$ is equal to the square of the ratio of the overlap factors of the first two echoes in the presence of a misalignment $\theta_1$ and a wedge angle $\alpha$.

\begin{figure} [htbp]
\centering\includegraphics[width=0.9\textwidth]{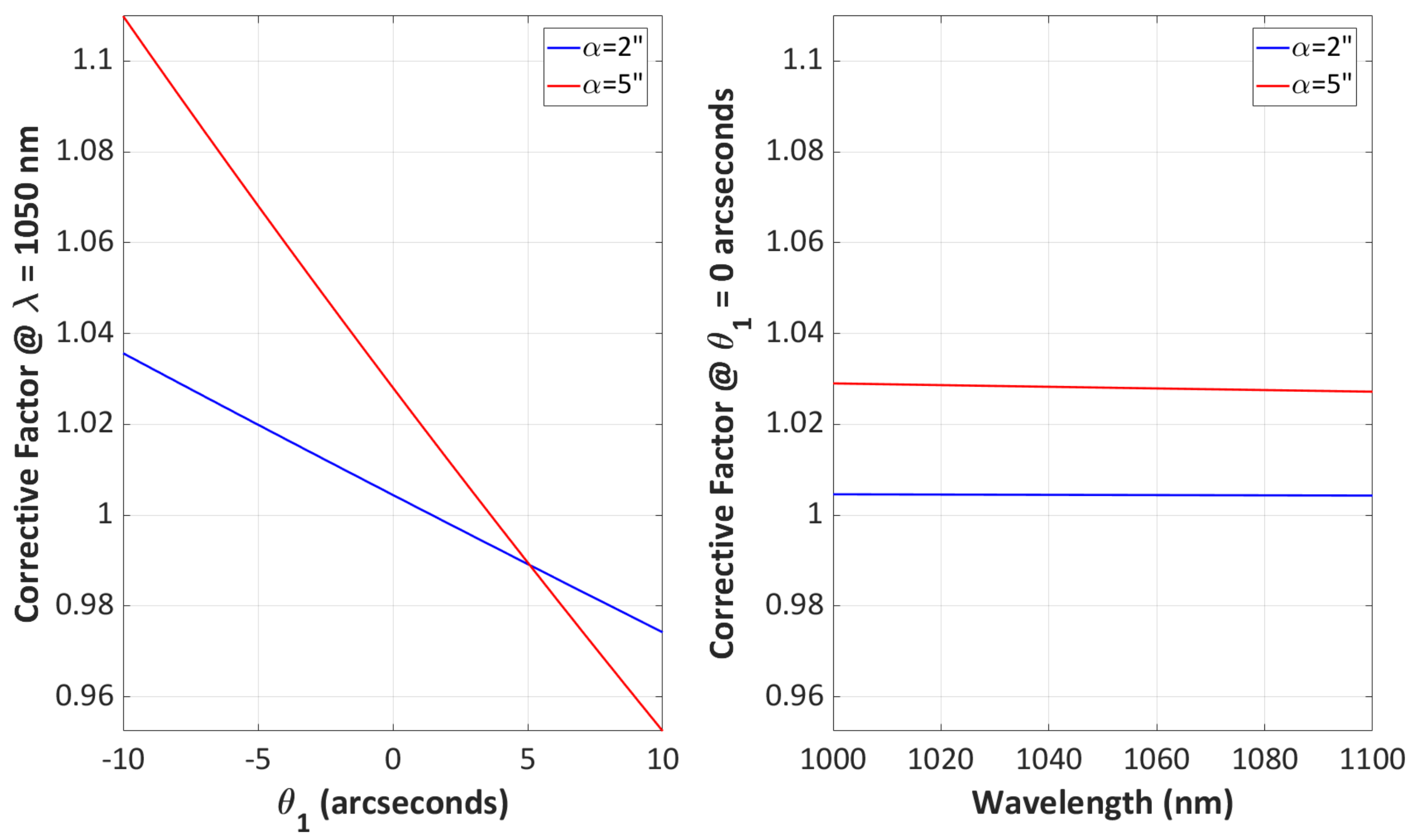}
\caption{Dependence of the corrective factor $\mathcal{K}$ on the angle of incidence $\theta_1$ (a) and the wavelength $\lambda$ (b) for wedge angle $\alpha$ equal to 2 arcseconds and 5 arcseconds.}
\label{fig:Corrective_Factor}
\end{figure}

Figure \ref{fig:Corrective_Factor} shows the dependence of this corrective factor $\mathcal{K}$ on the angle of incidence $\theta_1$ and the wavelength $\lambda$, for two wedge angles $\alpha$ equal to 2 arc seconds and 5 arc seconds respectively.
It can be seen that a very small wedge angle window ($\alpha\leqslant 2$ arcseconds) and a high alignment quality (better than $\pm5$ arcseconds) are needed to guarantee a relative measurement accuracy better than 1\%. Our alignment procedure implements an optimization of the recoupling of the SIG and REF beams in a single mode optical fiber placed at the image focus of a reflective collimator, the characteristics of the fiber and the collimator being identical to those of the items used in emission. This allows us to ensure that the error on the angle of incidence is comprised between 5 and 10 arc seconds. On the other hand, the spectral dependence of this corrective factor is extremely small under any circumstances, and can therefore be neglected.

\section{Experimental demonstration}
\label{sec:ExperimentalDemonstration}

\subsection{Operating conditions}
\label{sec:OperatingConditions}
In order to minimize measurement errors due to possible alignment bias, we have procured from \textsc{Light Machinery} \cite{Light_Machinery} a high quality test component, consisting of a 7980 A grade fused silica window of 2 mm thickness and 25 mm diameter, with a wedge angle $\alpha$ less than or equal to 2 arc seconds and one side of which is coated with a V-shaped anti-reflection coating centered at 1055 nm.

This component is installed in a piezoelectric gimbal (\textsc{Thorlabs} PGM1SE) comprising of two independently controllable rotational movements around the axes of the gimbal, that allow to maintain the center of the front face of the window in a fixed position when adjusting its angular orientation. The minimum adjustment step is 0.1 arc second for an angular range of approximately 1 degree.

The interferogram recordings were all made with a sampling rate of 100 kSamples/s and a digitizing range of $\pm 10$ Volts.

\subsection{Results}
\label{sec:Results}

\begin{figure} [htbp]
\centering\includegraphics[width=1.0\textwidth]{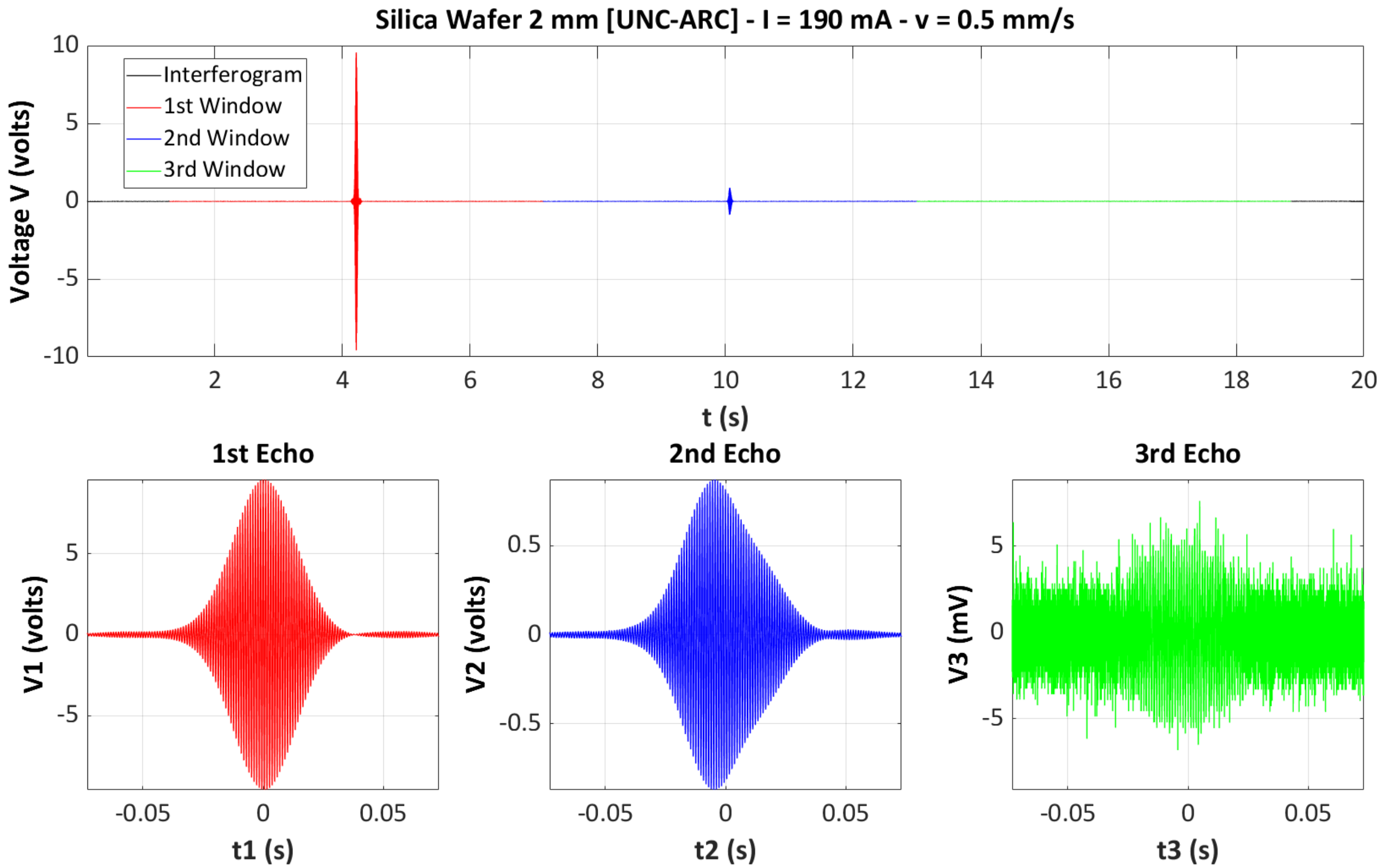}
\caption{Interferogram obtained from a 2 mm thick silica wafer with uncoated front side.}
\label{fig:UNC_ARC_190_0p5_data}
\end{figure}

The first interferogram (see Fig. \ref{fig:UNC_ARC_190_0p5_data}) is recorded for a window orientation where the front face is uncoated, a SLD driving current of 190 mA ($\lambda_0=1068$ nm, $\Delta\lambda = 33$ nm), and a translation speed $v$ of 0.5 mm/s.
The upper graph represents the whole recording (20 seconds, i.e. 2 Msamples), the black curve corresponding to the function $V(t)$ while the colored curves define the processing windows associated with the functions $V_m(t)$ for $m=1$ (in red), $m=2$ (in blue) and $m=3$ (in green). The lower graphs correspond to zoomed-in view of individual echoes, the voltage scale being adapted to the corresponding amplitude level of the echoes.

The portion of the interferogram before the first processing window can be used to estimate the variance of the voltage $V(t)$. We find: $\sigma_{V,\text{exp}}^2=1.93\times 10^{-6}\text{ Volts}^2$. This value must be compared to the one predicted by our theoretical approach [cf. section \ref{sec:DetectionNoise}, equations (\ref{eq:sigma2RIN}) and (\ref{eq:sigma2q})], namely
\begin{equation}
\sigma_{V,\text{th}}^2=\sigma_{V,q}^2+\sigma_{V,\text{RIN}}^2=2G^2\left\{2eSP_{\text{max}}+S^2\text{NEP}^2\right\}B+G^2 10^{(\text{RIN}-\text{CMRR})/10}S^2P_{\text{max}}^2B
\label{eq:sigma2Vth}
\end{equation}
Experimentally, the maximum power $P_{\text{max}}$, as expected, is defined by the digitizing range (0.45 mW). The theoretical contributions are thus distributed as follows
\begin{equation}
\sigma_{V,q}^2=3.02\times 10^{-7}\text{ Volts}^2\quad\text{;}\quad\sigma_{V,\text{RIN}}^2=5.12\times 10^{-(3+\text{CMRR}/10)}\text{ Volts}^2
\end{equation}
By comparing all these values, we can deduce a likely estimate of the CMRR of the NIRVANA receiver under our conditions of use, i.e. CMRR = 35 dB, which is in agreement with the manufacturer's data, the 50 dB of rejection being reached only in the autobalanced mode (which cannot be implemented in our case due to carrier frequency constraints).

As shown in Fig. \ref{fig:UNC_ARC_190_0p5_data}, we have chosen to take into account the third echo in order to be able to apply our method to the measurement of ultra-low reflection coefficients and thus to estimate its ultimate sensitivity. For this third echo, the equations (\ref{eq:FFTSquareModulusRatio}) and (\ref{eq:Rcoat}) become
\begin{equation}
\frac{|\mathcal{S}_3(F_l)|^2}{|\mathcal{S}_1(F_l)|^2}=\frac{|\rho_3(f_l)|^2}{|\rho_1(f_l)|^2}=\frac{|t_1(f_l)r_2(f_l)r'_1(f_l)r_2(f_l)t'_1(f_l)|^2}{|r_1(f_l)|^2}=T_1^2(f_l)R_2^2(f_l)
\label{eq:ThirdEcho_1}
\end{equation}
and
\begin{equation}
R_{\text{coat}}(f_l)=\frac{1}{1-R_s(f_l)}\frac{|\mathcal{S}_3(F_l)|}{|\mathcal{S}_1(F_l)|}
\label{eq:ThirdEcho_2}
\end{equation}

\begin{figure} [htbp]
\centering\includegraphics[width=1.0\textwidth]{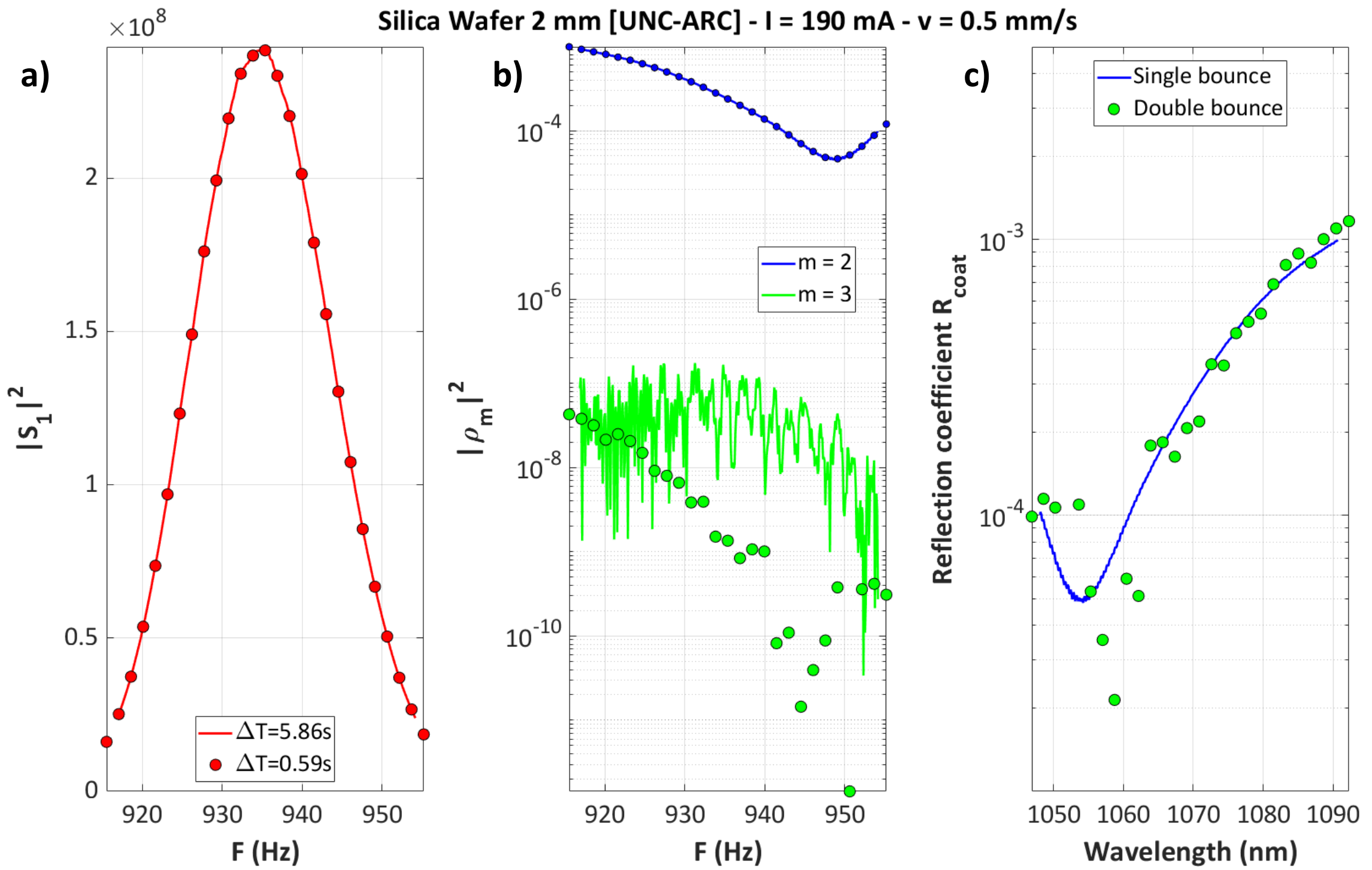}
\caption{Illustration of the different steps used in the data processing in the case of a 2 mm thick silica window coated with anti-reflection coating on the rear side (see text for more details).}
\label{fig:UNC_ARC_190_0p5_results}
\end{figure}

The graphs in Fig. \ref{fig:UNC_ARC_190_0p5_results} illustrate the different steps of our data processing. The graph a) shows the frequency dependence of the modulus squared of $\mathcal{S}_1(F)$, the discrete Fourier transform of the signal $V_1(t)$. The continuous red curve corresponds to the case where the width $\Delta T$ of the processing window is equal to the time interval separating two consecutive echoes, i.e. 5.86 s, while the red dots correspond to the result of the same DFT calculation, but for a window width 10 times smaller ($\Delta T=0.59$ s). This second curve is identical to the previous one, except that the frequency sampling pitch is 10 times larger.

The graph b) shows the frequency dependence of the coefficients $|\rho_2|^2$ (blue) and $|\rho_3|^2$ (green). The continuous curves are associated with the maximum time window width, while the discrete points are obtained with a window width reduced by a factor 10. The two processing modalities lead to identical results in the case of the second echo, while the correspondence is much less clear in the case of the third echo. This is simply due to the fact that the signal-to-noise ratio is much lower in this case, and that it is thus necessary to reduce the spectral resolution, by varying the width of the processing window, to obtain a better quality result.

This conclusion is perfectly confirmed by the curves gathered in the last graph (Fig. \ref{fig:UNC_ARC_190_0p5_results}c), which present the wavelength dependence of the reflection coefficient of the anti-reflective coated face. The continuous blue curve corresponds to the measurement result obtained from the second echo and presents a maximum spectral resolution (0.2 nm), while the green points correspond to the one obtained from the third echo with a lower resolution (2 nm). Note that we observe a very good agreement between these two independent determinations of the reflection coefficients above 100 ppm. It should be kept in mind that the signal levels used for the second determination are analogous to those that would be produced by the reflection on a coated face with a reflection coefficient between 0.1 and 40 ppb!

By introducing in equation (\ref{eq:rhomin_2}) the result of our experimental determination of the variance of the noise affecting the interferogram measurement, we can estimate the detection floor of our set-up in terms of measurement of the $|\rho_m|^2$ coefficients, namely
\begin{equation}
|\rho_m|_{\text{min}}^2\approx 10\left(\frac{\Delta\lambda}{\lambda_0}\right)^2\frac{\sigma_{V,\text{exp}}^2}{G^2S^2 P_{\text{max}}^2}\frac{F_0^2}{F_s}\Delta T=\left\{
\begin{aligned}
&7.3\times 10^{-10}\quad\text{for }\Delta T=5.86\text{ s}\\
&7.3\times 10^{-11}\quad\text{for }\Delta T=0.59\text{ s}
\end{aligned}
\right.
\label{eq:rho2_min_exp1}
\end{equation}
The second of these two values, where the signal level is sufficient for the comparison to be valid, is in satisfactory agreement with that which can be deduced from Fig. \ref{fig:UNC_ARC_190_0p5_results}b ($|\rho_3|_{\text{min}}^2\sim3\times 10^{-10}$).

\begin{figure} [htbp]
\centering\includegraphics[width=1.0\textwidth]{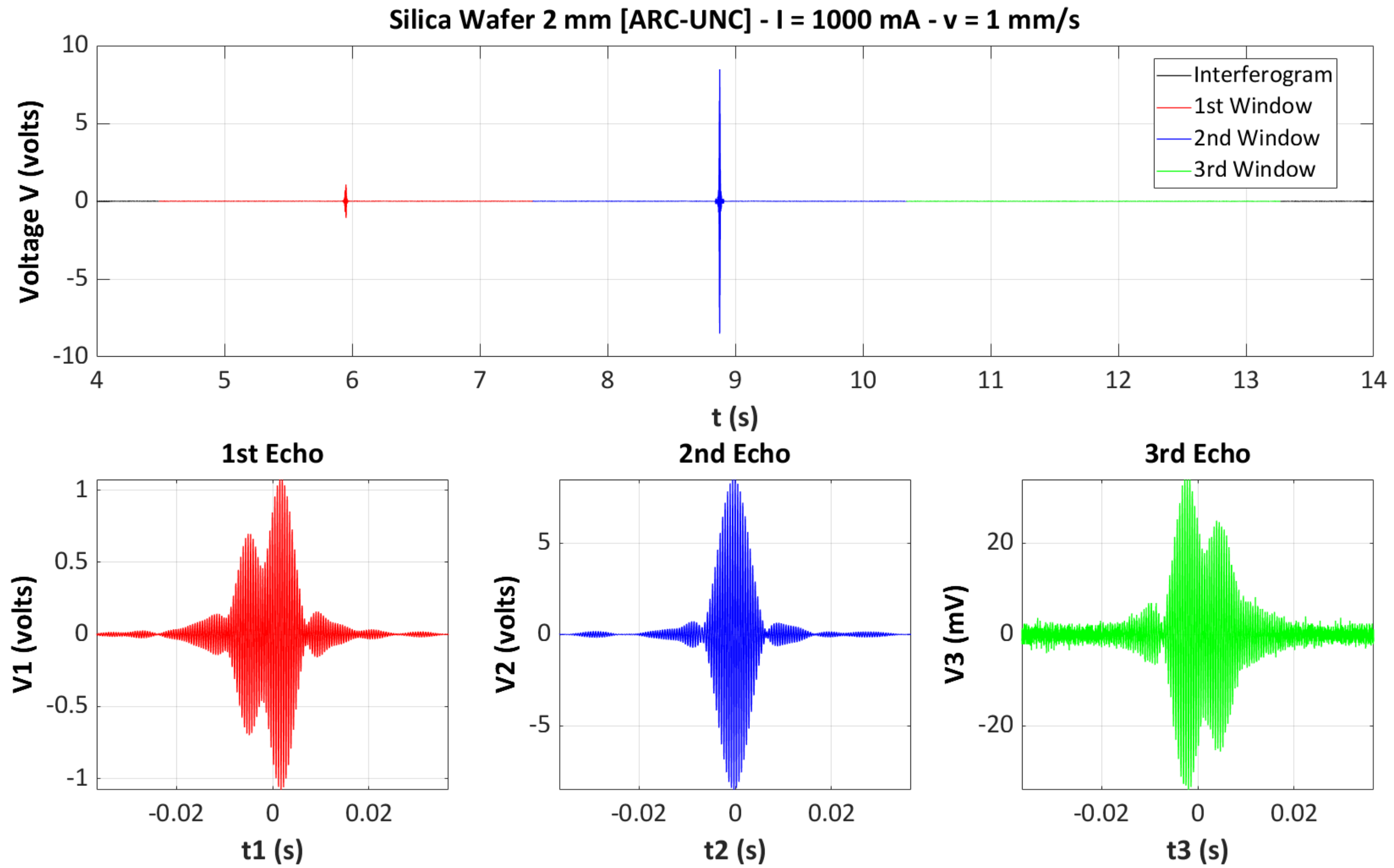}
\caption{Interferogram obtained on a 2 mm thick silica wafer with coated front side.}
\label{fig:ARC_UNC_1000_1p0_data}
\end{figure}

\begin{figure} [htbp]
\centering\includegraphics[width=1.0\textwidth]{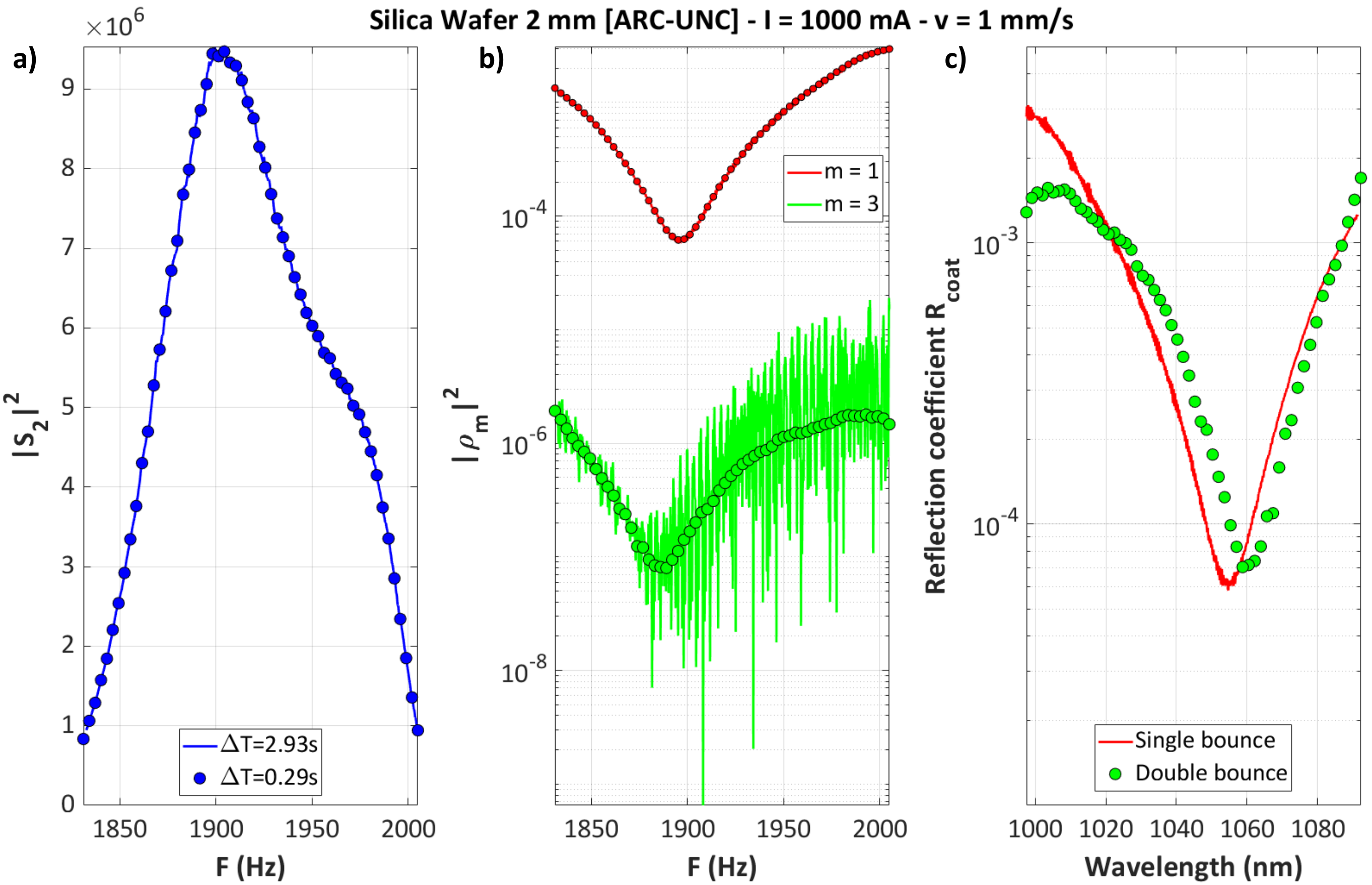}
\caption{Illustration of the different steps used in the data processing in the case of a 2 mm thick silica window coated with anti-reflection coating on the front side (see text for more details).}
\label{fig:ARC_UNC_1000_1p0_results}
\end{figure}

We note on Fig. \ref{fig:UNC_ARC_190_0p5_results}c that the V-shape characteristic of the behavior of this type of anti-reflection coating is only partially obtained, because the spectral range covered (1050 nm to 1090 nm, for the driving current used, i.e. 190 mA) does not allow its full measurement. Increasing the supply current to its maximum value (1000 mA) allowed us to cover a wider range of wavelengths (typically 1000 nm to 1090 nm), and thus to have access to the complete expected V-shape, for both possible orientations of the sample used. Figures \ref{fig:ARC_UNC_1000_1p0_data} and \ref{fig:ARC_UNC_1000_1p0_results} show the results obtained when the front side of the window corresponds to the coated side and the translation speed is 1 mm/s.
The highest amplitude echo is now the second one (the one corresponding to the reflection on the uncoated back side).

For this new window orientation, the equations (\ref{eq:ThirdEcho_1}) and (\ref{eq:ThirdEcho_2}) become (the calibration echo is indeed the second)
\begin{equation}
\frac{|\mathcal{S}_3(F_l)|^2}{|\mathcal{S}_2(F_l)|^2}=\frac{|\rho_3(f_l)|^2}{|\rho_2(f_l)|^2}=\frac{|t_1(f_l)r_2(f_l)r'_1(f_l)r_2(f_l)t'_1(f_l)|^2}{|t_1(f_l)r_2(f_l)t'_1(f_l)|^2}=R_1(f_l)R_2(f_l)
\end{equation}
and
\begin{equation}
R_{\text{coat}}(f_l)=\frac{1}{R_s(f_l)}\frac{|\mathcal{S}_3(F_l)|^2}{|\mathcal{S}_2(F_l)|^2}
\end{equation}

Fig. \ref{fig:ARC_UNC_1000_1p0_results}b shows the frequency dependence of the coefficients $|\rho_1|^2$ (red) and $|\rho_3|^2$ (green): as before, the continuous curves are associated with the maximum width of the processing time window, while the discrete points are obtained with a window width reduced by a factor of 10. The two processing modalities lead this time to identical results in the case of the two echoes. Indeed, the measurement on the third echo is now identical to the one that would be produced by the reflection on a coated face with a reflection coefficient between 80 ppb and 1.6 ppm (two reflections on the uncoated face and one reflection on the coated face), to be compared to the previous case (0.1 ppb and 40 ppb, for two reflections on the coated face and one reflection on the uncoated face): the signal-to-noise ratio is therefore much better.

Using in equation (\ref{eq:rho2_min_exp1}) the parameters corresponding to this second experimental test [$\lambda_0=1042$ nm, $\Delta\lambda=82$ nm, $F_0=1.92$ kHz, $\Delta T=2.93$ s], the minimum value of the coefficients $|\rho_m|^2$ that can be measured with a signal to noise ratio of 10 is
\begin{equation}
|\rho_m|_{\text{min}}^2\approx\left\{
\begin{aligned}
&10^{-8}\quad\text{for }\Delta T=2.93\text{ s}\\
&10^{-9}\quad\text{for }\Delta T=0.29\text{ s}
\end{aligned}
\right.
\label{eq:rho2_min_exp2}
\end{equation}
Again, these theoretical estimations are consistent with the experimental results obtained with this reversed window orientation.

The agreement between the two experimental determinations of the reflection coefficient of the coated face presented on Fig. \ref{fig:ARC_UNC_1000_1p0_results}c is very satisfactory in terms of overall shape, but reveals the presence of a slight spectral shift between the two curves (on the order of 5 nm).

\begin{figure} [htbp]
\centering\includegraphics[width=0.65\textwidth]{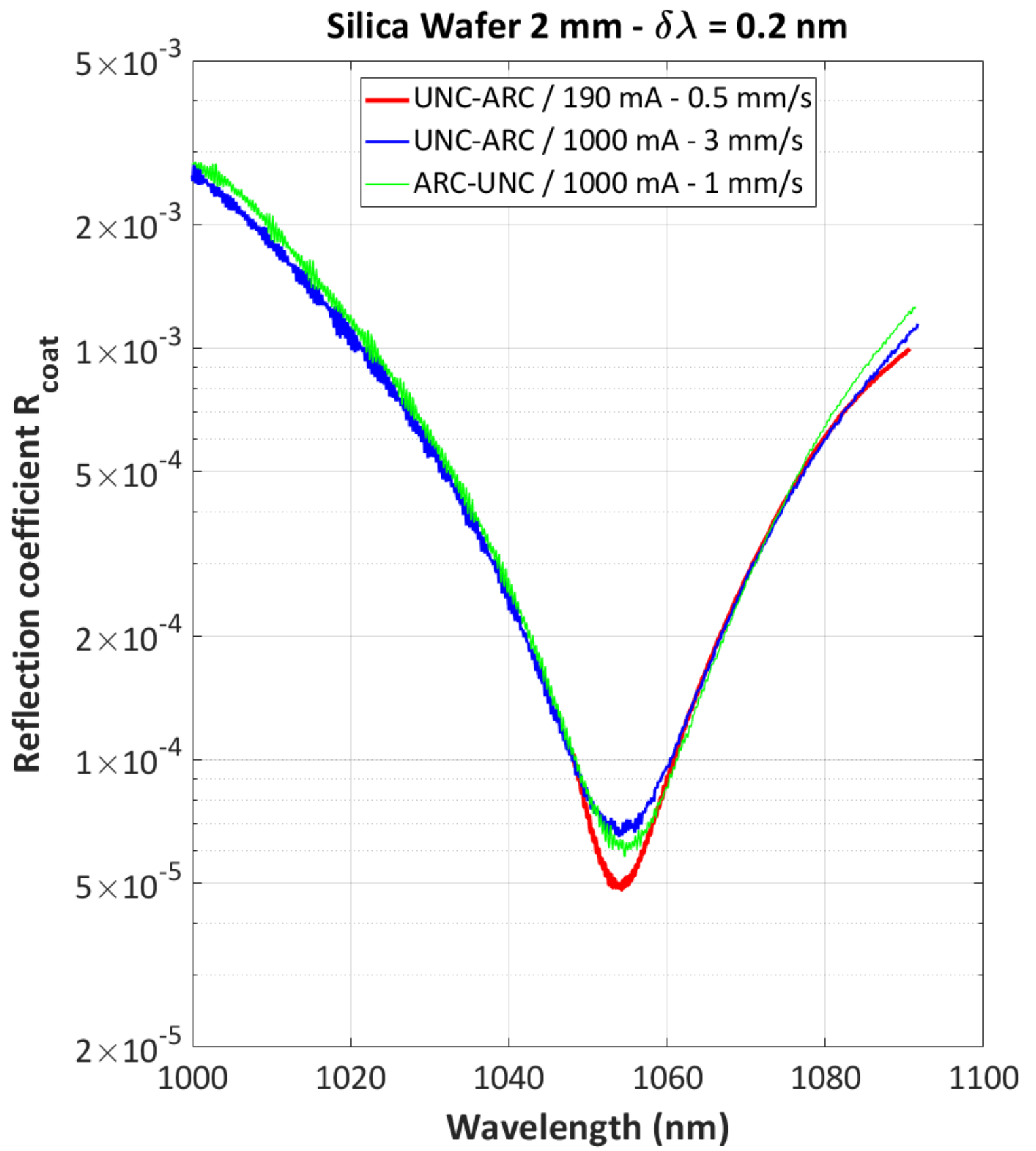}
\caption{Comparative presentation of the measured reflection coefficient of the coated face obtained under different experimental conditions.}
\label{fig:Results_Synthesis}
\end{figure}

Before proposing possible explanations for this experimental observation in Section \ref{sec:Discussion}, it seems important to present in a comparative way all the measurements we have made on this sample in a simple bounce configuration (see Fig. \ref{fig:Results_Synthesis}).
Note that the results obtained are not only independent of the translation speed of the hollow retro-reflector and the driving current of the superluminescent diode, but also of the orientation of the window (UNC for uncoated, ARC for anti-reflection coated).

Let us now turn to the phase measurements that is possible thanks our processing method. In accordance with the description of this method made in Section \ref{sec:DataProcessing}, the phase shift at the reflection on the coated face is equal to the opposite of the argument of the discrete Fourier transform (DFT) of the signal associated with the first echo, when the front face of the window corresponds to its coated face [see equation (\ref{eq:Phase})].

\begin{figure} [htbp]
\centering\includegraphics[width=0.95\textwidth]{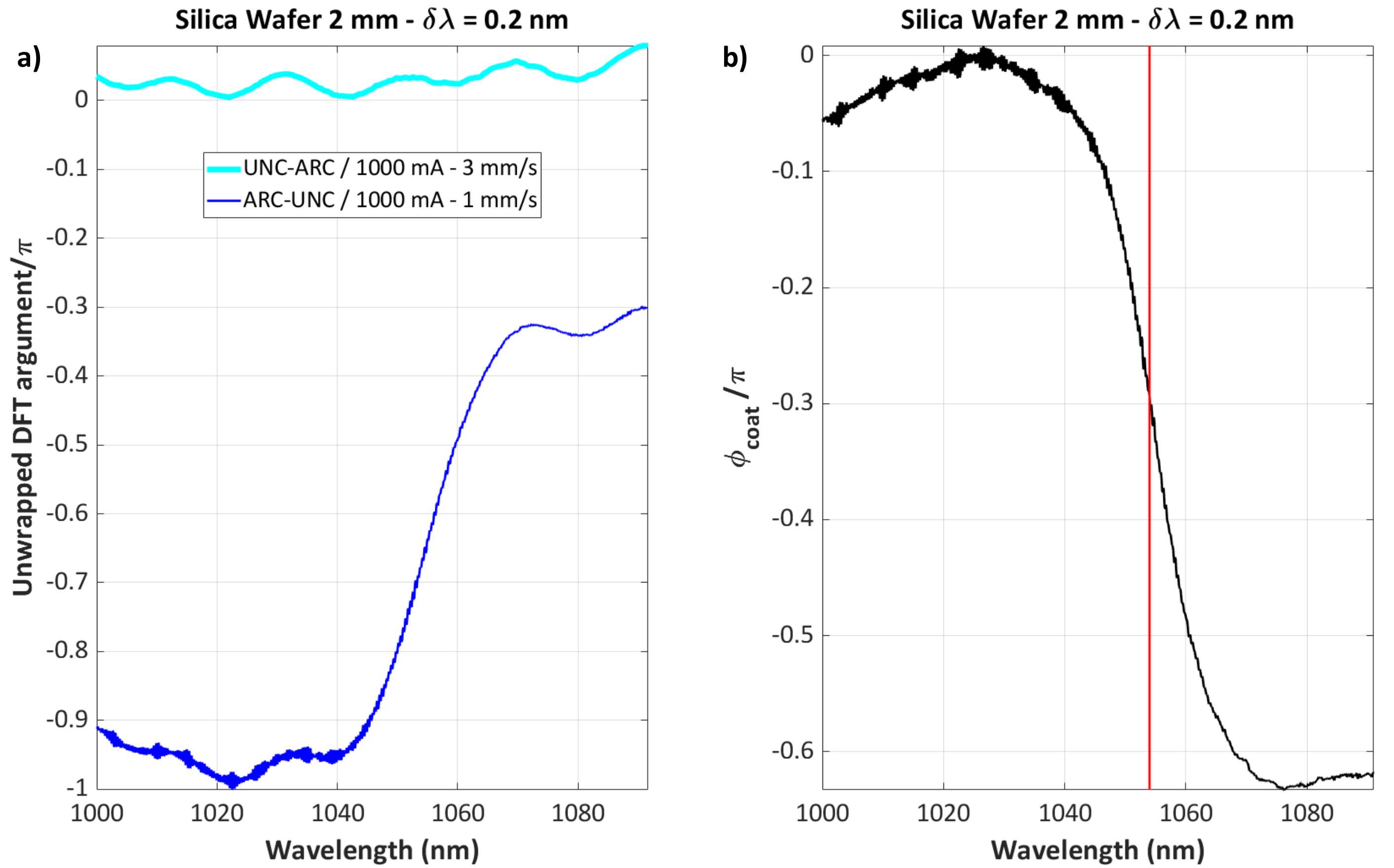}
\caption{a) spectral dependence of the unwrapped DFT argument for the two window orientations (light blue, uncoated face in front; dark blue, coated face in front) - b) spectral dependence of the phase change at the reflection on the coated face.}
\label{fig:Phase}
\end{figure}

The experimental result is presented in Fig. \ref{fig:Phase}a (dark blue curve).
The general appearance of the spectral dependence of this phase corresponds to what we would expect in the case of a V-coating, but presents some small oscillations, whose origin is difficult to determine since the exact layers stack formula is not known to us. We have therefore chosen to apply the same experimental approach to the case where the front face of the sample corresponds to its uncoated face. Indeed, in this case, the theoretical result is known (phase shift equal to $\pi$ on the whole spectral range). The experimentally obtained curve corresponds to the light blue curve of Fig. \ref{fig:Phase}a. It is approximately constant (average value of about 0.05$\pi$), but also presents the same small oscillations as those observed in the case where the front face of the sample corresponds to its coated face. A possible solution to this problem consists of transposing to this phase measurement the calibration principle used during the determination of the $|\rho_m|^2$ coefficients, but this time taking into account the two orientations of the test window. Under these conditions, the equation (\ref{eq:Phase}) becomes
\begin{equation}
\text{arg}[r_{\text{coat}}(f_l)]-\text{arg}[r_{\text{uncoat}}(f_l)]=-\text{arg}[\mathcal{S}_{1,\text{coat}}(F_l)]+\text{arg}[\mathcal{S}_{1,\text{uncoat}}(F_l)]
\end{equation}
or, taking into account the value of the phase shift on the uncoated side
\begin{equation}
\phi_{\text{coat}}(f_l)=\pi+\text{arg}[\mathcal{S}_{1,\text{uncoat}}(F_l)]-\text{arg}[\mathcal{S}_{1,\text{coat}}(F_l)]
\label{eq:PhaseDifference}
\end{equation}
The result of this calibration procedure is shown in \ref{fig:Phase}b (black curve). The shape of the spectral dependence thus obtained is much more satisfactory, with in particular the existence of a very clear symmetry around the wavelength corresponding to the coating reflection minimum $\lambda_{\text{min}}$ (marked by a vertical red line on Fig. \ref{fig:Phase}b).

\section{Discussion}
\label{sec:Discussion}

The experimental results presented in Section \ref{sec:Results} are in very good agreement with the predictions of our theoretical model. The only two small deviations that we could note are:
\begin{itemize}
\item[\textbullet] the wavelength shift between the reflection spectra of the coated face obtained in single bounce and double bounce for a 1000 mA driving current (see Fig. \ref{fig:ARC_UNC_1000_1p0_results}),
\item[\textbullet] the necessity to implement a calibration procedure in the case of the measurement of the phase shift induced by the reflection on the coated face.
\end{itemize}
The objective of the following sections is to identify possible explanations for these two discrepancies.

\subsection{Wedge angle}
\label{sec:WedgeAngle}

As pointed out in Section \ref{sec:AlignementBias}, the presence of a wedge angle on the substrate causes a lateral shift between the beam reflected from its rear side and that reflected from its front side. Obviously, the same phenomenon occurs between the double bounce and the single bounce inside the window, the beam deviation associated with the third echo being defined by
\begin{equation}
\beta_3=2\theta_1-4n_s\alpha=\beta_1-4n_s\alpha=\beta_2-2n_s\alpha
\end{equation}
However, the modeling result presented in Section \ref{sec:AlignementBias} shows that the spectral dependence of this effect is negligible, and remains so in the case of the differential shift between third and second echo. Furthermore, the results obtained with a driving current of 190 mA (see Fig. \ref{fig:UNC_ARC_190_0p5_results}c) show that this effect is not always present and is therefore not solely attributable to the wedge angle between the window faces.

\subsection{Far field filtering}
\label{sec:FarFieldFiltering}

Even if it is unlikely that the spectral shift is related to it, it seemed useful to us to test the influence, on the measurement result, of the geometric filtering in the far field carried out by the detection photodiodes. To do this, we replaced the Nirvana receiver by a receiver \textsc{Thorlabs} PDB210C in which the diameter $2a$ of the active area of InGaAs photodiodes is 3 mm. At the same time, the bandwidth $B$ of the RF output is increased to 1 MHz, while the CMRR is simply 30 dB, for a trans-impedance gain $G$ of $5\times 10^{5}$ V/A and a NEP of 16 pW/$\sqrt{\text{Hz}}$. 

Using the equations (\ref{eq:Eta_a}) and (\ref{eq:wd}), we deduce from these functional characteristics the new value of the geometric overlap coefficient, i.e. $\eta_a =0.995$, which confirms that far-field filtering is indeed removed.

As in the case of the Nirvana balanced receiver, the maximum power is defined by the condition (\ref{eq:PmaxADC}) on the digitizing range
\begin{equation}
P\leqslant\frac{10}{2\sqrt{2}\eta_aT_{\text{ref},1}GS\sqrt{R_s}}\sim 0.3\text{ mW}
\end{equation}
We deduce the value of $P_{\text{max}}$
\begin{equation}
P_{\text{max}}=\eta_aT_{\text{ref},1}P=44\thinspace\mu\text{W}
\end{equation}
then that of the variance of voltage fluctuations using equation (\ref{eq:sigma2Vth})
\begin{equation}
\sigma_{V,\text{th}}^2=\sigma_{V,q}^2+\sigma_{V,\text{RIN}}^2=8.75\times 10^{-5}\text{ Volts}^2+9.8\times 10^{-6}\text{ Volts}^2=9.73\times 10^{-5}\text{ Volts}^2
\end{equation}

\begin{figure} [htbp]
\centering\includegraphics[width=1.0\textwidth]{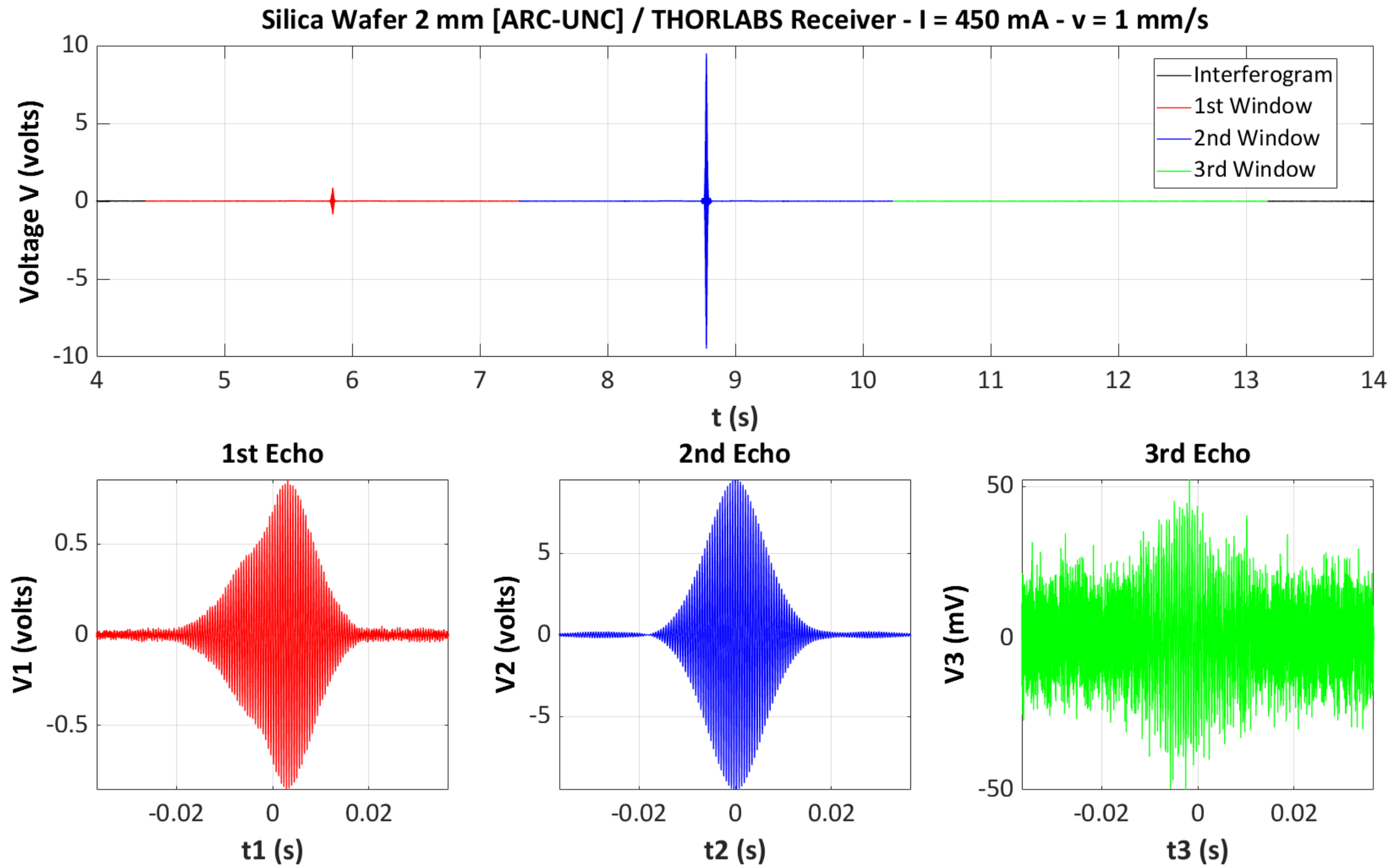}
\caption{Interferogram obtained with the PDB210C THORLABS receiver on the 2mm thick silica window coated on one side (driving current SLD 450 mA, translation speed 1 mm/s).}
\label{fig:ARC_UNC_Scan_THORLABS}
\end{figure}

\begin{figure} [htbp]
\centering\includegraphics[width=1.0\textwidth]{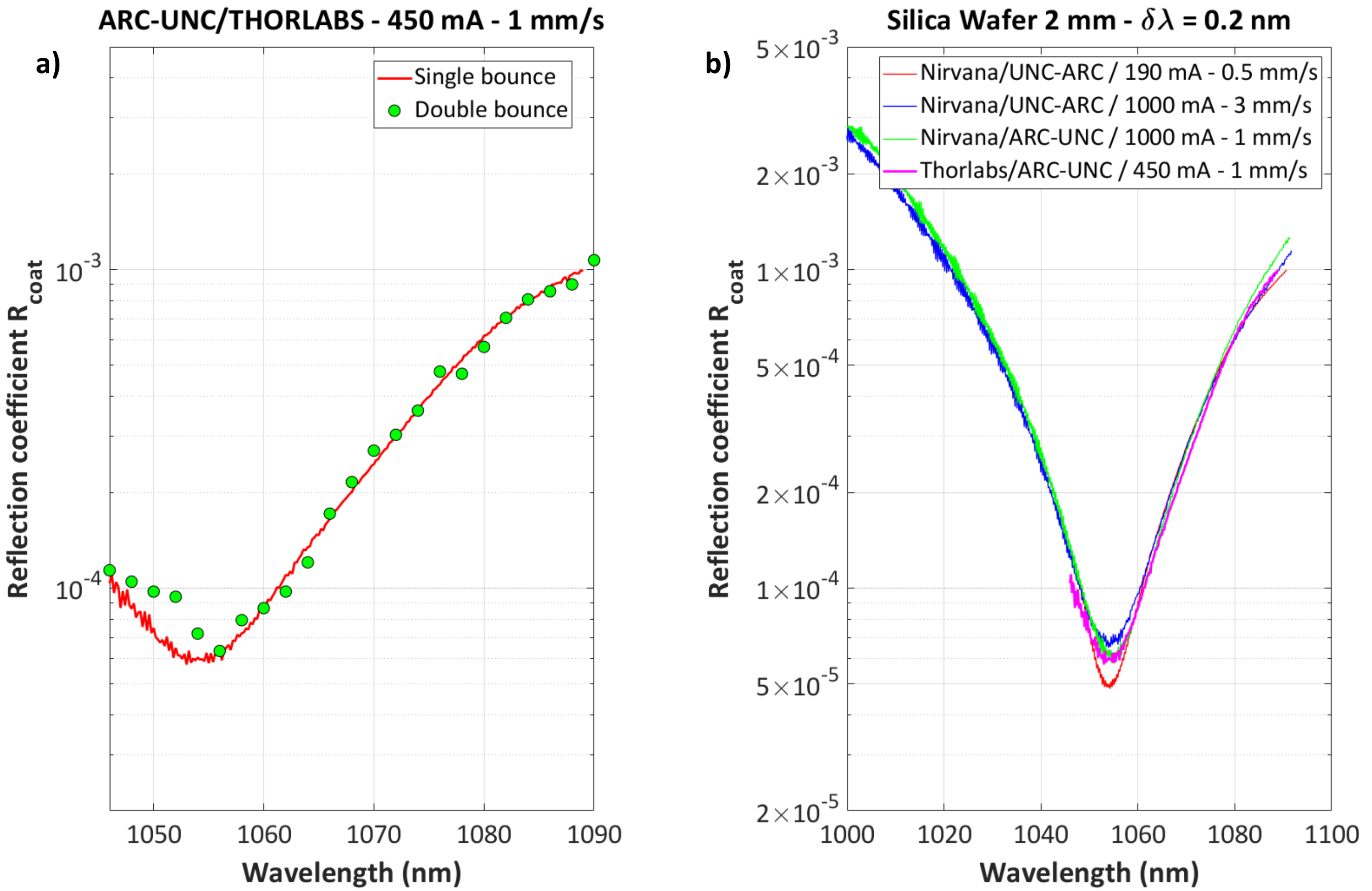}
\caption{Comparative presentation of the reflection spectra of the coated face obtained (a) with the Thorlabs PDB210C receiver in single and double bounce configuration - (b) with the Nirvana receiver (red, blue and green curves) and the Thorlabs PDB210C receiver (magenta curve).}
\label{fig:Thorlabs}
\end{figure}

Figure \ref{fig:ARC_UNC_Scan_THORLABS} shows the interferogram obtained with the PDB210C \textsc{Thorlabs} receiver on our test sample when the front side of the silica window corresponds to its coated side (SLD driving current 450 mA, i.e. $\lambda_0=1060$ nm, $\Delta\lambda_0=44$ nm; translation speed 1 mm/s).
As mentioned in section \ref{sec:Results}, the portion of the interferogram before the first processing window can be used to estimate the variance of the voltage $V(t)$. We get: $\sigma_{V,\text{exp}}^2=1.05\times 10^{-4}\text{ Volts}^2$, which is in very good agreement with our theoretical prediction. This result allows us to estimate the value of the smallest coefficient $|\rho_m|^2$ that can be detected with a signal-to-noise ratio of 10 using this set-up. It comes
\begin{equation}
|\rho_m|_{\text{min}}^2\approx 10\left(\frac{\Delta\lambda}{\lambda_0}\right)^2\frac{\sigma_{V,\text{exp}}^2}{G^2S^2 P_{\text{max}}^2}\frac{F_0^2}{F_s}\Delta T=\left\{
\begin{aligned}
&6\times 10^{-7}\quad\text{for }\Delta T=2.93\text{ s}\\
&6\times 10^{-8}\quad\text{for }\Delta T=0.29\text{ s}
\end{aligned}
\right.
\label{eq:rho2_min_Thorlabs}
\end{equation}
This detection floor is significantly higher than what we obtained with the Nirvana receiver [see equations (\ref{eq:rho2_min_exp1}) and (\ref{eq:rho2_min_exp2})] and may make it difficult to detect a possible shift between the reflection spectra obtained in single bounce or double bounce configuration. To try to circumvent this problem, we decided to carry out, under the same experimental conditions, 4 successive recordings of the same interferogram and to average the wavelength dependence of the reflection coefficients thus obtained. The result of this approach is shown in Fig. \ref{fig:Thorlabs}a.
It is quite clear, in particular for values of the reflection coefficient higher than 100 ppm, where the signal-to-noise ratio is sufficient, that these two curves do not show any detectable shift (one will usefully refer to Figure \ref{fig:ARC_UNC_1000_1p0_results}c to better perceive the difference in behavior).

Moreover, this change of detector has confirmed that the modeling of the signal-to-noise ratio that we presented in Section \ref{sec:DetectionNoise} accurately describes the influence of the various characteristic parameters of the balanced receiver on the quality of the measurement. And at the end, Fig. \ref{fig:Thorlabs}b shows that the results we obtain with this method do not depend on the receiver used, which is again very satisfactory and guarantees the reproducibility of these measurements.

\subsection{Dispersive OPD}
\label{sec:DispersiveOPD}

To establish equation (\ref{eq:V}), we explicitly assumed that the optical path difference $\Delta$ did not depend on the frequency $f$. This is a reasonable assumption, as the configuration has been defined to approximate this condition \cite{Khan_2021}. Let us explain our approach further: from the separation of the two channels by the splitter cube BS$_1$, each of them travels the following paths in a glass
\begin{equation}
\text{SIG: }c_1/2+c_1+c_2/2\quad\text{;}\quad\text{REF: }c_1/2+p+c_2/2
\label{eq:OPDCompensation}
\end{equation}
where $c_j$ is the size of the splitter cube BS$_j$, ($j=1,2$) and $p$ is the size of the right angle prism RAP.

The equation (\ref{eq:OPDCompensation}) shows that this right angle prism plays the same role as a compensating plate in a Michelson interferometer. But, this compensation is only realized if $p=c_1$. The optical components used are made of N-BK7 ($n=1.505@1064$ nm) with a size of 25.4 mm and a manufacturing tolerance of $\pm 0.25$ mm. This means that the OPD compensation is realized with an uncertainty of $\delta\Delta=(n-1)(p-e_1)\sim\pm250$ $\mu$m. 

We must therefore take into account the possible presence of a compensation error in our set-up. This will induce a spectral dependence of the position of the zero OPD and thus make the shape of the interferogram associated with the first echo asymmetric. When this compensation is perfectly realized in a Fourier transform spectrometer, because of this symmetry, it is not necessary to record the interferogram for positive or negative values of the optical path difference \cite{Fellgett_1958}.

\begin{figure} [htbp]
\centering\includegraphics[width=0.95\textwidth]{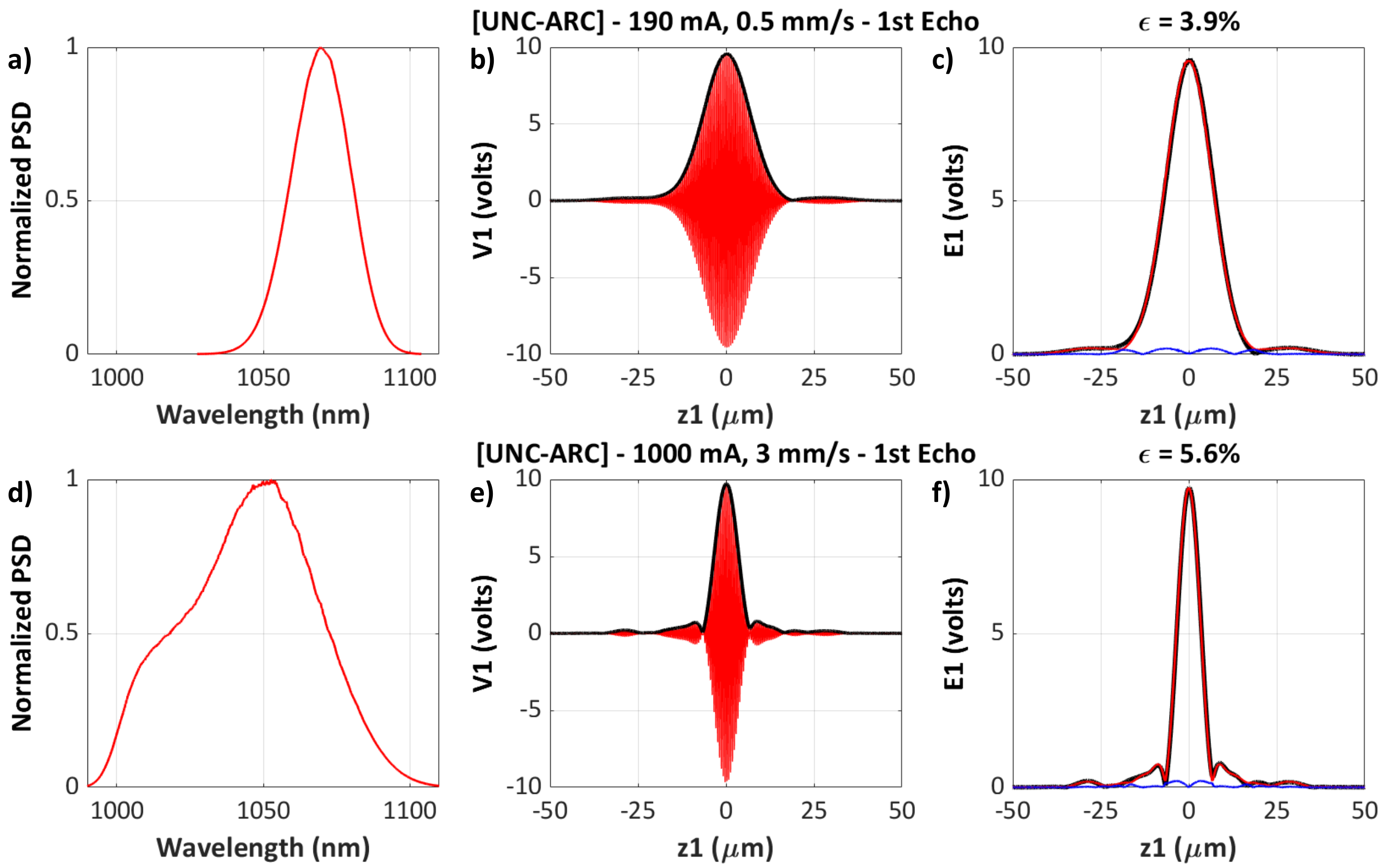}
\caption{Analysis of the asymmetric character of the interferograms recorded when reflection is from the uncoated face of the test sample [a) and d) normalized power spectral density - b) and e) variation of the voltage $V_1$ as a function of the displacement $z_1$ of the translation stage (red curve) and envelope of this signal (black curve) - c) and f) decomposition of the envelope (black curve) into even (red curve) and odd (blue curve) parts ; a), b), and c), driving current of 190 mA - d), e), and f), driving current of 1000 mA].}
\label{fig:Asymetric}
\end{figure}

First of all, we will analyze whether the first echoes recorded from the uncoated side of our test sample are symmetrical or not, and quantify their possible asymmetry. To this end, we start by calculating the signal envelope of recorded voltage $V_1$ using a Hilbert transform \cite{Pavlicek_2021,Pikalek_2014,Xin_2020}, namely
\begin{equation}
\mathcal{E}_1(t)=\mathcal{H}\{V_1\}(t)=\frac{1}{\pi}\text{PV}\int\limits_{-\infty}^{+\infty}\frac{V_1(\tau)}{t-\tau}d\tau
\end{equation}
where PV denotes the Cauchy principal value. Then we decompose this envelope into even and odd parts
\begin{equation}
\mathcal{E}_1(t)=\frac{\mathcal{E}_1(t)+\mathcal{E}_1(-t)}{2}+\frac{\mathcal{E}_1(t)-\mathcal{E}_1(-t)}{2}=\mathcal{E}_{1,e}(t)+\mathcal{E}_{1,o}(t)
\end{equation}
and quantify the asymmetry using the following $\epsilon$ quantity
\begin{equation}
\epsilon=\frac{\displaystyle\int\limits_{-\infty}^{+\infty}|\mathcal{E}_{1,o}(t)|\thinspace dt}{\displaystyle\int\limits_{-\infty}^{+\infty}\mathcal{E}_{1,e}(t)\thinspace dt}
\end{equation}
Figure \ref{fig:Asymetric} shows the result of this analysis for two different driving currents, namely 190 mA and 1000 mA.
The asymmetry of the interferogram varies between 3.9\% for a driving current of 190 mA and 5.6\% for a driving current of 1000 mA. This clearly confirms the presence of a weakly dispersive part in the optical path difference of our set-up.

To determine the influence of such a dispersive part, we have developed a simulation program of our set-up under MATLAB and introduced in the expression of the optical path of the REF channel, the crossing of a thickness $e$ of N-BK7. The expression of the optical path difference then becomes
\begin{equation}
\Delta(f)=2vt+e[n_{\text{BK7}}(f)-1]
\end{equation}

\begin{figure} [htbp]
\centering\includegraphics[width=0.9\textwidth]{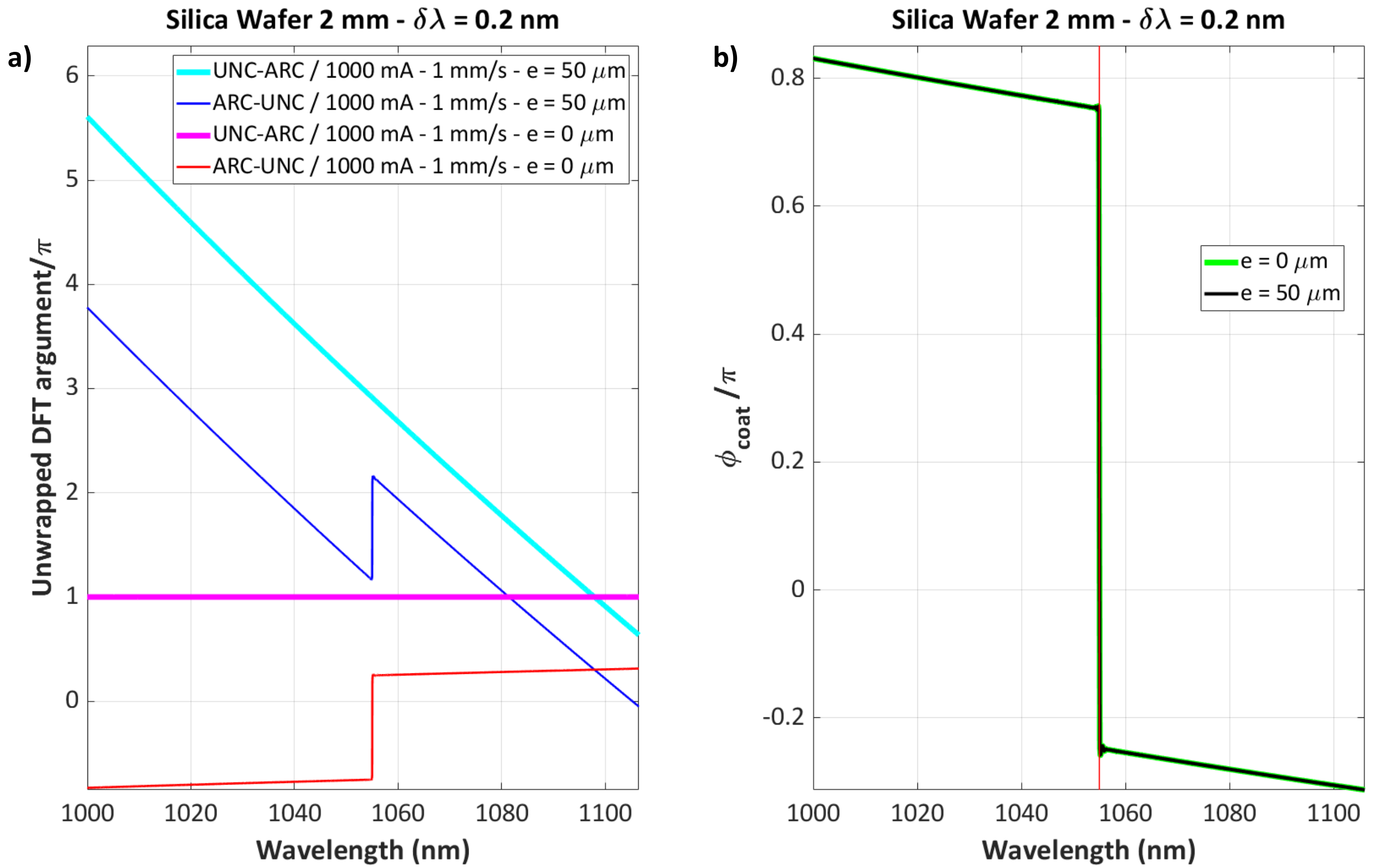}
\caption{Influence of the spectral dispersion of the optical path difference on the measurement result of the reflection phase shift on an anti-reflection coated interface - a) Spectral dependence of the phase of the discrete Fourier transform of the echo associated with the uncoated face (UNC-ARC, magenta curve in the absence of dispersion, cyan curve in the presence of dispersion) and that associated with the coated face (ARC-UNC, red curve in the absence of dispersion, blue curve in the presence of dispersion) ; b) Spectral dependence of the phase shift on the reflection obtained by using the calibration procedure described in the text (green curve in the absence of dispersion, black curve in the presence of dispersion, the vertical red line indicating the centering wavelength of the anti-reflection coating).}
\label{fig:Dispersion}
\end{figure}
This numerical modeling shows that the presence of such a dispersion of optical path difference has no influence on the result of the reflection coefficient measurements. Figure \ref{fig:Dispersion} shows the impact of this dispersion on the results of a phase measurement, when the anti-reflection coating is a SiO$_2$/Nb$_2$O$_5$ bilayer deposited on a fused silica substrate and centered at 1050 nm. On graph a), the magenta and red curves are relative to the case without dispersion for the two possible orientations of the sample, respectively UNC-ARC and ARC-UNC. The cyan and blue curves are relative to the dispersive case (N-BK7 thickness of 50 microns) for these two orientations. The presence of a dispersive optical path difference causes the appearance of a linear variation of phase as a function of the wavelength, but the implementation of the calibration procedure defined at the end of the section \ref{sec:Results} allows to find perfectly the information of phase shift sought (see Fig. \ref{fig:Dispersion}b). This validates and justifies the implementation of such a procedure. On the other hand, the presence of a dispersive OPD does not explain the slight oscillations appearing on our experimental results. Further analysis will therefore be necessary to understand their origin and to give our phase measurement results the ultimate metrological quality.

\section{Conclusion}
\label{sec:Conclusion}

In this paper, we have demonstrated that the implementation of a numerical processing transposed from that used in Fourier transform spectrometry to retro-reflection data acquired by a low-coherence balanced detection interferometer scanned in optical path difference allows the determination of the spectral dependence of the reflection coefficient of V-shaped anti-reflection coatings over a spectral range on the order of 100 nm, with a spectral resolution of up to 0.2 nm and a detection floor on the order of 0.1 ppm. Moreover, a posteriori choice of the width of the processing windows allows to improve this performance by a factor 10 (10 ppb) at the cost of a correlated degradation of the spectral resolution (2 nm).

Our method also allows the determination of the spectral dependence of the phase shift due to the reflection on a coated interface. However, further tests seem necessary to quantify the possible influence of a residual spectral dispersion of the optical path difference of the interferometer on the metrological quality of this phase measurement.

The modeling of the key parameters of the balanced receiver and their influence on the signal-to-noise ratio using this new measurement method has been experimentally verified, in particular by using detection systems from two different manufacturers.

The lowest coefficient of reflection that can be measured with our set-up is defined by equation (\ref{eq:rhomin_2}), in which two quantities are implicit functions of $v$, the translation speed, namely
\begin{equation}
F_0=\frac{2v}{\lambda_0}\quad\text{and}\quad\Delta T=\frac{n_gd_s}{v}
\end{equation}
Accordingly, equation (\ref{eq:rhomin_2}) can be written in the following equivalent form
\begin{equation}
|\rho_m(f_l)|_q^2=10\left(\frac{\Delta\lambda}{\lambda_0}\right)^2\frac{2eB}{SP_{\text{max}}}\frac{4v}{F_s}\frac{n_gd_s}{\lambda_0^2}
\label{eq:rhomin_3}
\end{equation}
This shows that an optimized choice of two of the key parameters of the balanced receiver (saturation power of the photodiodes $P_{\text{max}}$ and detection bandwidth $B$), as well as the modification of some of the experimental conditions (decrease of the translation speed $v$ down to 0.1 mm/s, increase of the sampling frequency $F_s$ up to 1 MHz) should allow to significantly improve the sensitivity of our setup. Moreover, this increase in saturation power will require the use of larger diameter photodiodes, which will offer more flexibility and tolerance in the alignment procedure. The application of all these modifications should result in an overall gain of 3 decades and thus would pave the way for the use of this new measurement method for the characterization of light scattered by optical interfaces, coated or not.

\section*{Acknowledgments}
This work is part of the StrayLight Working Group for the Laser Instrument Group of the LISA Consortium. The authors would like to thank the AMIdex Talents Management Program, Aix Marseille University and the French National Space Center (CNES) for their financial support.

\section*{Disclosures}
The authors declare no conflicts of interest.

\section*{Data availability}
Data underlying the results presented in this paper are not publicly available at this time but may be obtained from the authors upon reasonable request.


\end{document}